\documentclass[prx,amsfonts,floatfix,superscriptaddress,twocolumn,longbibliography]{revtex4-1}
\usepackage{mathbbol}
\usepackage{graphicx,color}
\usepackage{amsmath,amssymb,bm,amsthm}
\usepackage{mathtools}
\usepackage{simplewick}
\usepackage{braket}
\usepackage{epstopdf}

\def\be{\begin{equation}}
\def\ee{\end{equation}}
\def\ba{\begin{eqnarray}}
\def\ea{\end{eqnarray}}

\begin{document}
\title{Thouless and relaxation time scales in many-body quantum systems}

\author{Mauro Schiulaz}
\address{Department of Physics, Yeshiva University, New York City, NY, 10016, USA}
\author{E. Jonathan Torres-Herrera}
\address{Instituto de F\'isica, Benem\'erita Universidad Aut\'onoma de Puebla,
Apt. Postal J-48, Puebla, 72570, Mexico}
\author{Lea F. Santos}
\address{Department of Physics, Yeshiva University, New York City, NY, 10016, USA}

\date{\today}

\begin{abstract}
A major open question in studies of nonequilibrium quantum dynamics is the identification of the time scales involved in the relaxation process of isolated quantum systems that have many interacting particles. We demonstrate that long time scales can be analytically found by analyzing dynamical manifestations of spectral correlations. Using this approach, we show that the Thouless time, $t_{\text{Th}}$, and the relaxation time, $t_{\text{R}}$, increase exponentially with system size. We define $t_{\text{Th}}$ as the time at which the spread of the initial state in the many-body Hilbert space is complete and verify that it agrees with the inverse of the Thouless energy. $t_{\text{Th}}$ marks the point beyond which the dynamics acquire universal features, while relaxation happens later when the evolution reaches a stationary state. In chaotic systems, $t_{\text{Th}}\ll t_{\text{R}}$, while for systems approaching a many-body localized phase, $t_{\text{Th}}\rightarrow t_{\text{R}}$.  Our analytical results for $t_{\text{Th}}$ and $t_{\text{R}}$ are obtained for the survival probability, which is a global quantity. We show numerically that the same time scales appear also in the evolution of the spin autocorrelation function, which is an experimental local observable. Our studies are carried out for realistic many-body quantum models. The results are compared with those for random matrices. 
\end{abstract}

\maketitle

\section{Introduction}

There is currently great interest in the dynamics of isolated interacting many-body quantum systems. This is in part due to the advances of experiments with cold atoms, ion traps, and nuclear magnetic resonance platforms, which allow for the simulation of unitary dynamics of highly tunable Hamiltonians for long times~\cite{Gavish2005,Bloch2008,Gadway2010,Bloch2012,Jurcevic2014,Richerme2014,Schreiber2015,Garttner2017,Wei2018}. Great efforts have been devoted to conciliate reversible microscopic dynamics and irreversible thermodynamics~\cite{Gogolin2016,Borgonovi2016,Dalessio2016,DymarskyARXIV,Reimann2018a,*Reimann2018b}. Increasing attention has also focused on the analysis of the metal-insulator transition~\cite{Santos2005,Nandkishore2015,Luitz2017,Serbyn2015,*Serbyn2017,Varma2017} and the quantum-classical correspondence, especially in the context of many-body quantum chaos and the scrambling of quantum information~\cite{ScaffidiARXIV,Rammensee2018,Cotler2017GUE,Gharibyan2018,Nosaka2018,Chan2018,Borgonovi2019,BorgonoviARXIV}. A missing piece in these studies is a complete picture of the time scales involved in the relaxation to equilibrium.

Several works have discussed what equilibration in closed finite quantum systems actually means~\cite{Peres1984,Deutsch1991,Srednicki1996,Reimann2008,Short2011,Short2012,HeSantos2013,Zangara2013},  a subject on which we find consensus.  Equilibration refers to the proximity of an observable to its asymptotic value for most times, despite the presence of temporal fluctuations. Much more problematic is the identification of the time to reach equilibrium, for which there are several interesting, but contradictory results. Some suggest that equilibration happens at very short times, while others indicate just the opposite, that extremely long times are required~\cite{Monnai2013,Goldstein2013,Malabarba2014,Goldstein2015,Gogolin2016,Reimann2016,Pintos2017,Oliveira2018,DymarskyARXIVThouless}. 

To properly determine the relaxation time of many-body quantum systems, one needs to have a complete picture of the different behaviors that emerge at different time scales. Without that, one risks reaching misleading conclusions. Here, we unveil the time scales by using an analytical expression that describes the entire evolution of the survival probability for chaotic interacting systems. The survival probability is the squared overlap between the initial state and its time evolved counterpart. The crucial observation needed to obtain our analytic expression is that chaotic systems have strongly correlated eigenvalues that show level statistics comparable to what one finds for full random matrices~\cite{Guhr1998,MehtaBook}.

An expression for the evolution of the survival probability was proposed in Ref.~\cite{Torres2018} for a disordered spin-1/2 model in the chaotic regime. Here, we present all the steps involved in the analytical derivation, which is not tied to any specific model. The only assumptions made are that the system is defined on a finite lattice, has local two-body interactions only, is strongly chaotic, and that its initial state is far from equilibrium and has energy away from the borders of the spectrum. We confirm the generality of our equation by showing that it describes the whole evolution of the survival probability for the following chaotic models: a disordered  spin-1/2 model, a clean (dynamical) spin-1/2 model, and a sparse banded random matrix model.

In hands of the analytical equation for the survival probability, we arrive at one of the central results of this work: analytical estimates for two long timescales. One is what we call Thouless time, $t_\textrm{Th}$, which is the time for the survival probability to reach its minimum value at the bottom of the correlation hole, and the other is the relaxation time, $t_\textrm{R}$, which happens later, when the survival probability saturates to an asymptotic value. The correlation hole is a dip below the asymptotic value~\cite{Leviandier1986,Guhr1990,Wilkie1991,Alhassid1992,Gorin2002}, that has been observed in local many-body Hamiltonians with~\cite{Torres2017,Torres2017Philo} and without disorder~\cite{Torres2017Philo} and in the Sachdev-Ye-Kitaev model~\cite{Cotler2017GUE,Gharibyan2018,Nosaka2018}. 

The Thouless time was first introduced in the context of noninteracting systems, where it refers to the timescale for a particle to diffuse through a disordered metallic sample and reach the boundaries~\cite{Thouless1974}. This definition has been shown to agree with the inverse of the Thouless energy $E_\textrm{Th}$, which is the energy scale below which universality holds~\cite{Altshuler1986,*Altshuler1988}. Our studies bring to light the fact that these two approaches -- diffusion and spectral correlations -- give different results for interacting systems. We demonstrate that our definition of the Thouless time is indeed inversely proportional to the Thouless energy generalized to interacting systems in~\cite{Bertrand2016}. In contrast, our  $t_{\text{Th}}$ does not agree with definitions that employ transport properties~\cite{Serbyn2017,Varma2017,DymarskyARXIV,DymarskyARXIVThouless}.

According to our physical interpretation, the Thouless time in interacting systems refers to the time that it takes for an initially localized many-body state to fully spread in the exponentially large many-body Hilbert space accessible to its energy. This picture explains why the two approaches used to define the Thouless time in noninteracting systems are not equivalent for interacting ones. For single particle models, the Hilbert space coincides with the physical space, so complete spread in the former implies complete spread in the latter. The situation is quite different for many-body systems, for which the dimension $D$ of the Hilbert space is exponentially large in the physical size. Complete spread in the many-body Hilbert space requires a time exponentially large in the system size. 

We find that the Thouless time depends on the size of the Hilbert space as $t_{\text{Th}} \propto D^{2/3}/\Gamma$, where $\Gamma$ is the width of the energy distribution of the initial state. The relaxation time, $t_{\text{R}}\propto D/\Gamma$, also extracted directly from our analytical equation for the survival probability, coincides with the Heisenberg time, which is the longest possible timescale for the system. Both scalings are confirmed by exact numerical simulations.

These results are compared with the timescales obtained analytically for full random matrices from a Gaussian Orthogonal Ensemble (GOE). While the expression for the relaxation time still coincides with the Heisenberg time, full spreading in the Hilbert space is reached at a time which is independent of the matrix size.

In addition to the survival probability, which is a global quantity, we also investigate the local spin autocorrelation function, which is equivalent to the density imbalance measured in experiments with cold atoms~\cite{Schreiber2015}. Using a disordered spin-1/2 model, we show that the timescales for the spin autocorrelation function to reach the minimum of the correlation hole and to later saturate coincide with those found for the survival probability. 

A natural question that emerges from these studies is what happens to the timescales outside the chaotic region. To address this point, we investigate the dynamics of the disordered spin model as the disorder strength increases and the model leaves the chaotic regime toward a many-body localized phase, where the eigenvalues are no longer correlated. This affects the dynamics before~\cite{Torres2015} and after the Thouless time~\cite{Torres2017,Torres2017Philo}.  
We show that $t_{\text{Th}}$ grows exponentially with the disorder strength and approaches the relaxation time, that is $t_{\text{R}}/t_{\text{Th}} \rightarrow 1$. In noninteracting systems, this ratio is known as Thouless dimensionless conductance.

The remainder of this article is organized as follows. In Sec.~\ref{sec:definitions}, we provide the general structure of the models considered and introduce the survival probability. In Sec.~\ref{sec:GOE}, we study numerically and analytically the timescales for the survival probability evolving under GOE Hamiltonians. In Sec.~\ref{Sec:Realistic}, we present the analytical equation for the survival probability in realistic chaotic interacting models and use it to obtain $t_{\text{Th}}$ and $ t_{\text{R}}$ analytically. The expression is compared with numerical results for three realistic models of various system sizes. 
In Sec.~\ref{sec:numerics}, we study numerically how the timescales for the disordered spin-1/2 model change as the system approaches localization in space. We also show that our definition for the Thouless time is inversely proportional to the Thouless energy. In Sec.~\ref{sec:I}, we study numerically the spin autocorrelation function and find that the long timescales agree with those for the survival probability. In Sec.~\ref{sec:conclusion}, we summarize our results 
and outline some future directions. Appendix~\ref{sec:derivation} describes the steps involved in the derivation of the expression for the survival probability for realistic chaotic systems.

\section{General definitions}
\label{sec:definitions}

The systems studied in this article are described by real and symmetric Hamiltonians of the form
\be
H=H_0 + g V.
\label{eq:mu}
\ee
We take $\hbar =1$. $H_0$ is the integrable part of $H$, $V$ represents the perturbation, and $g=1$ is the perturbation strength.  The eigenvalues and eigenstates of $H$ are labeled $E_{\alpha}$ and $|\psi_{\alpha}\rangle$, respectively.

The system is prepared in an eigenstate $|  \Psi(0) \rangle$ of $H_0$ with energy 
\be
E_{0} =  \langle \Psi(0) |H|  \Psi(0) \rangle
\ee
close to the middle of the spectrum. Due to $V$, the initial state spreads in time in the many-body basis defined by $H_0$. The perturbation takes the system very far from equilibrium. To study the evolution of the initial state, we compute the survival probability
\be
P_S (t)=\left| \langle \Psi(0) | e^{-iHt} | \Psi(0) \rangle \right|^2,
\label{Eq:SP}
\ee
which represents the probability to find the system in the initial state at time $t$.

The survival probability allows for two different integral representations. The first one is obtained by writing it as
\ba
P_S(t)&=& \left| \sum_{\alpha} \left| C_\alpha^{(0)} \right|^2 e^{-i E_\alpha t} \right|^2
\!\!=\! \left| \int \rho_{0}(E) e^{-iEt} dE \right|^2 \!\!,
\label{Eq:intPS} 
\ea
where
 $C_\alpha^{(0)}=\left< \psi_{\alpha} |\Psi(0)\right>$ is the component of the initial state over the energy eigenbasis and
\be
\rho_{0}(E)=\sum_{\alpha} \left| C_\alpha^{(0)} \right|^2 \delta(E-E_{\alpha})
\label{Eq:LDOS}
\ee
is the energy distribution of the initial state, which is also known as local density of states (LDOS) or strength function. The width $\Gamma$ of this distribution is given by
\be
\Gamma^2 = \sum_{n \neq 0} |  \langle \phi_n |H| \Psi(0) \rangle |^2,
\label{eq:Gamma}
\ee
where $|\phi_n \rangle $ are the eigenstates of $H_0$. $\Gamma^2$ is related to the number of states $|\phi_n \rangle $ directly coupled to the initial state by $V$.

We take averages over initial states with energies close to the middle of the spectrum, $E_0 \sim 0$. For random models, we also average over different realizations of the Hamiltonian. We denote the total average by $\langle . \rangle$. For clean models, the average is performed only over initial states.

The asymptotic value of the survival probability corresponds to its infinite time-average,
\begin{equation}
\overline{P_S} =\left<\sum_{\alpha}\left|C_\alpha^{(0)}\right|^4\right>.
\label{SPtimeave_SM}
\end{equation}
If the coefficients $C_\alpha^{(0)}$ are Gaussian random numbers satisfying normalization, $\overline{P_S} \sim 3/D$, where $D$ is the size of the many-body Hilbert space.

\section{Time scales for the survival probability in the GOE model}  
\label{sec:GOE}

The first model that we study corresponds to GOE random matrices. We take $H_0$ to be the diagonal part of the random matrix $H$ and $V$ to be the off-diagonal part. The elements are independent random numbers from a Gaussian distribution with mean 0 and variance 2 for $H_0$ and 1 for $V$. The model is unrealistic, since it implies the simultaneous interaction between all particles, but it allows for the identification of universal properties.

For matrices with a large dimension $D$, the analytical expression for the entire evolution of the survival probability under GOE matrices is given by~\cite{Torres2018,SantosTorres2017AIP}
\begin{equation}
\left<P_S(t)\right>=\frac{1-\overline{P_S}}{D-1}\left[D\frac{\mathcal{J}_1^2(2 \Gamma t)}{(\Gamma t)^2}-b_2\left(\frac{\Gamma t}{2D}\right)\right]+\overline{P_S},
\label{Eq:PsGOE}
\end{equation}
where ${\cal J}_1(t)$ is the Bessel function of the first kind,  the two-level form factor is
\ba
b_2(t) &=& [1-2t + t \ln(1+2 t)] \Theta (1- t) \label{eq:b2} \\
&+& \{t \ln [ (2 t+1)/(2 t -1) ] -1 \} \Theta(t-1),\nonumber
\ea
and $\Theta$ is the Heaviside step function. Following Eq.~(\ref{eq:Gamma}), $\Gamma = \sqrt{D} $ for the GOE model.
 
\begin{figure}[htb]
\includegraphics[scale=0.35]{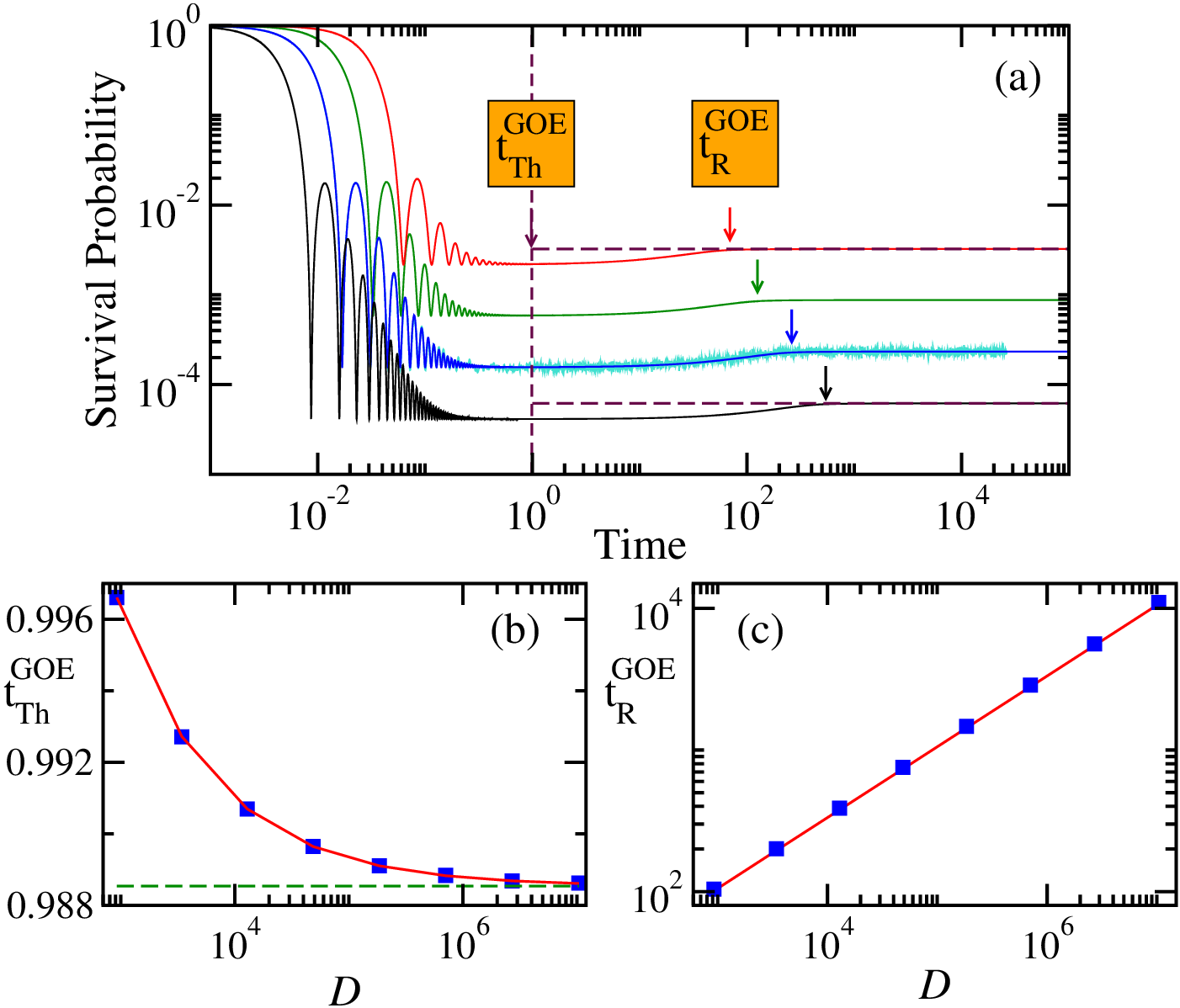}
\caption{Survival probability for the GOE model. (a) Analytical expression for the survival probability as a function of time [Eq.~(\ref{Eq:PsGOE})] for GOE matrices of dimensions $D=924, 3\,432, 12\,870,  48\,620$, from top to bottom. For $D=12\,870$, we also provide the numerical curve. The time scales $t_\textrm{Th}^{\text{GOE}}$ and $t_\textrm{R}^{\text{GOE}}$ are marked for each curve. (b) The time $t_\textrm{Th}^{\text{GOE}}$ to reach the minimum of the correlation hole as a function of $D$. The data converge to the asymptotic value $(3/\pi)^{1/4}$ of Eq.~(\ref{Eq:tthGOE}) (horizontal dashed line) as $1/\sqrt{D}$ (solid line). (c) Relaxation time $t_\textrm{R}^{\text{GOE}}$ as a function of $D$. The data follow the behavior $t_\textrm{R} \simeq (1/3)\sqrt{D/\delta}$ (solid line) obtained in Eq.~(\ref{Eq:tRGOE}).  }
\label{fig:GOE}
\end{figure}

A plot of the analytical Eq.~(\ref{Eq:PsGOE}) is provided in Fig~\ref{fig:GOE}~(a) for different sizes $D$ of the Hamiltonian matrix. The numerical curve for $D=12870$ is also shown and, apart from fluctuations at long times, it is undistinguishable from the analytical expression. The evolution of $\left<P_S(t)\right>$ is initially determined by $\mathcal{J}_1^2(2 \Gamma t)/(\Gamma t)^2$, which at very short times gives $1 - \Gamma^2 t^2$ and later leads to oscillations that follow a power-law decay $\propto t^{-3}$. This decay persists until the minimum of $\left<P_S(t)\right>$ is reached at a time that we call $t_{\text{Th}}^{\text{GOE}}$.
After  $t_{\text{Th}}^{\text{GOE}}$, $\left<P_S(t)\right>$ is dominated by the $b_2(t)$ function and increases toward saturation. The $b_2(t)$ function describes the correlation hole. This dip below the saturation point is a direct manifestation of the rigidity of the spectrum, being nonexistent in integrable models, where the level spacing distribution is Poissonian. 

\subsection{Time for the Minimum of the Correlation Hole}

We use Eq.~(\ref{Eq:PsGOE}) to compute the dependence of  $t^{\textrm{GOE}}_\textrm{Th}$ on  $D$. Since the first term in Eq.~(\ref{Eq:PsGOE}) depends on $\Gamma t$, while the second term depends on $\Gamma t/D$, we expect the minimum of $\left<P_S(t)\right>$ to happen at times which are large with respect to $1/\Gamma \sim 1/ \sqrt{D}$, but short with respect to $D/\Gamma\sim \sqrt{D}$. As a consequence, we expand the first term of Eq.~(\ref{Eq:PsGOE}) for long times,
\begin{equation}
D\frac{\mathcal{J}_1^2(2 \Gamma t)}{(\Gamma t)^2} \rightarrow  \frac{D}{\pi (\Gamma t)^3} \hspace{0.4 cm } \text{for} \hspace{0.4 cm } \Gamma t\gg1,
\label{Eq:expandJ}
\end{equation}
and expand the two-level form factor $b_2$ for short times,
\begin{equation}
b_2\left(\frac{\Gamma t}{2D}\right) \rightarrow 1 -\frac{\Gamma t}{D} \hspace{0.4 cm } \text{for} \hspace{0.4 cm } \frac{\Gamma t}{D}\ll1.
\label{Eq:expandb}
\end{equation}
Combining Eq.~(\ref{Eq:expandJ}) and Eq.~(\ref{Eq:expandb}) in the derivative of $\left<P_S(t)\right>$, we have
\begin{equation}
\left.\frac{d\left<P_S(t)\right>}{dt}\right|_{t=t_{\text{Th}}^{\text{GOE}}} \!\!\! \simeq \left.  \frac{1-\overline{P_S}}{D-1}\left[
-3 \frac{D}{\pi \Gamma^3 t^4} +
 \frac{\Gamma }{D} 
\right]   \right|_{t=t_{\text{Th}}^{\text{GOE}}}  \!\!\!= 0.
\end{equation}
In the fully connected GOE model, all factors that depend on $D$ cancel out, resulting in
\be
t_{\text{Th}}^{\text{GOE}}=  \left(\frac{3}{\pi} \right)^{1/4} \frac{\sqrt{D}}{\Gamma} = \left(\frac{3}{\pi} \right)^{1/4} .
\label{Eq:tthGOE}
\ee
While the initial decay determined by $\Gamma$ gets faster with $D$, the subsequent power-law decay lasts for longer, which leads to the constant value of $t_{\text{Th}}^{\text{GOE}}$. This is in stark contrast with physical chaotic models, where, as we shall see in Sec.~\ref{Sec:Realistic}, $t_{\text{Th}} $ grows with system size.

The minimum value reached by the survival probability can be found by plugging Eq.~(\ref{Eq:tthGOE}) into Eq.~(\ref{Eq:PsGOE}), which gives
\ba
\left. \left<P_S(t)\right> \right|_{t=t_{\text{Th}}^{\textrm{GOE}}} &\approx& \frac{1-\overline{P_S}}{D-1}\left[\frac{D}{\pi (\Gamma t_{\textrm{Th}}^{\textrm{GOE}} )^3}  - \left(1 -\frac{\Gamma t_{\textrm{Th}}^{\textrm{GOE}} }{D} \right)\right]\nonumber\\
 &+& \overline{P_S}\sim \frac{1-\overline{P_S}}{D-1} (-1) +\overline{P_S}.
\ea
Since all eigenstates of GOE matrices are Gaussian random vectors, so is $|\Psi(0)\rangle$. This implies that $\overline{P_S}\sim3/D$ and
\be
\left. \left<P_S(t)\right> \right|_{t=t_{\text{Th}}^{\textrm{GOE}}} \approx \frac{2}{D}.
\ee

It is worth comparing our result in Eq.~(\ref{Eq:tthGOE})  with Ref.~\cite{Alhassid1992}, where the expression for $\left<P_S(t)\right>$ does not properly capture the short time decay. As a consequence, it is found there, incorrectly, that $t_{\text{Th}}^{\text{GOE}}$ scales with $D$. If, however, the matrix elements are rescaled by a factor $1/\sqrt{D}$, as done in~\cite{Cotler2017GUE}, so that the width of the density of states is independent of $D$,  then Eq.~(\ref{Eq:tthGOE}) changes and $t_{\text{Th}}^{\text{GOE}}$ becomes indeed dependent on $D$.

In Fig.~\ref{fig:GOE} (b) we plot the dependence of $t_\textrm{Th}^{\text{GOE}}$ on $D$. The data are obtained by numerically minimizing Eq.~(\ref{Eq:PsGOE}). As we can see, $t_\textrm{Th}^{\text{GOE}}$ converges asymptotically to the value given in Eq.~(\ref{Eq:tthGOE}), which is indicated with the horizontal dashed line. A power-law fitting of the data gives $0.25/\sqrt{D}$, which is shown with the solid line.

\subsection{Relaxation Time} 
\label{sec:tR}

To estimate the relaxation time, we study the relative difference between $\left<P_S(t)\right>$ and $\overline{P_S}$. To do so, we expand the two-level form factor for long times:
\begin{equation}
b_2\left(\frac{\Gamma t}{2D}\right) \rightarrow  \frac{D^2}{3\Gamma^2 t^2} \hspace{0.4 cm } \text{for} \hspace{0.4 cm } \frac{\Gamma t}{D}\gg1.
\label{Eq:longB2}
\end{equation}
We also neglect the term involving the Bessel function, since it goes to zero faster than quadratically for $t\rightarrow \infty$. Substituting Eq.~(\ref{Eq:longB2}) into Eq.~(\ref{Eq:PsGOE}) gives  
\begin{equation}
\frac{\left|\left<P_S(t)\right>-\overline{P_S}\right|}{\overline{P_S}} \approx \frac{1-\overline{P_S}}{\overline{P_S}(D-1)} \frac{D^2}{3\Gamma^2 t^2} \approx \left( \frac{D}{3\Gamma t} \right)^2.
\end{equation}
This shows that $\left<P_S(t)\right>$ approaches the saturation value following a power-law behavior, so the timescale for complete relaxation is not well defined. Yet, one can define the relaxation time as the point where
\begin{equation}
\frac{\left|\left<P_S(t_{\text{R}})\right>-\overline{P_S}\right|}{\overline{P_S}} \sim \delta ,
\label{eq:delta}
\end{equation}
for some small value $\delta>0$. This gives
\be
t_{\text{R}}^\textrm{GOE}\sim \frac{D}{3\Gamma\sqrt{\delta}}\sim\frac{1}{3}\sqrt{\frac{D}{\delta}}.
\label{Eq:tRGOE}
\ee

The relaxation time is therefore inversely proportional to the mean level spacing $\Gamma/D$, which is the definition of the Heisenberg time. This is the largest possible timescale for a quantum system, derived directly from Eq.~(\ref{Eq:PsGOE}). Unlike $t_{\text{Th}}^{\text{GOE}}$, the time to reach actual saturation diverges with $D$.

As for $\delta$, we choose a value $\delta\ll\sigma_{P_S}/\overline{P_S}$, where $\sigma_{P_S}$ is the width of the ensemble fluctuations of $P_S$ at asymptotically long times. Since $\sigma_{P_S}\sim \overline{P_S}$~\cite{TorresKollmar2015}, this implies $\delta\ll 1$. In our plots we take $\delta=0.01$.

In Fig~\ref{fig:GOE} (c), we plot the dependence of $t_R^\textrm{GOE}$ on $D$. The numerical data (squares) are compared with the analytical prediction of Eq.~(\ref{Eq:tRGOE}), finding perfect agreement. No fitting parameters were used for this comparison.

\section{Timescales for the survival probability in realistic chaotic models} 
\label{Sec:Realistic}

The GOE model is not appropriate to describe physically relevant chaotic systems. This is so because, in a random matrix model, no notion of locality is present and simultaneous interaction of all degrees of freedom is assumed. As a consequence, one cannot expect a priori the predictions of Sec.~\ref{sec:GOE} to hold for realistic models.

In this section, we provide an analytical equation for $\left<P_S(t)\right>$ for generic chaotic many-body quantum systems. With this analytical expression, we find estimates for the timescales for the evolution of the survival probability. These predictions are then checked against numerical data. We find that, while the behavior of the system at short times is very different from that of the GOE, at long times the two models behave in an equivalent way. This is because the dynamics at long times depend on spectral correlations only.

\subsection{Analytical expression for the survival probability}
\label{sec:Analytical}
We consider a many-body quantum system on a lattice, in the strongly chaotic regime.  The interactions are local and two body only, which implies that the density of states has a Gaussian shape~\cite{Brody1981}.  In the bulk of the spectrum, the eigenstates of these systems are close to Gaussian random vectors. 

When the system is taken very far from equilibrium, as done here [in Eq.~(\ref{eq:mu}), $g = 1$], initial states with $E_{0} \sim 0$ are very delocalized in the energy eigenbasis~\cite{Torres2014PRA,Torres2014NJP}. In this case, the LDOS defined in Eq.~(\ref{Eq:LDOS}) is also Gaussian. 
Because the coupling determined by the $V$ part of the total Hamiltonian is local and short range, $H$ is a very sparse matrix and, according to Eq.~(\ref{eq:Gamma}), the width $\Gamma$ of the LDOS is proportional to $\sqrt{L} \ll \sqrt{D}$. This is a main difference from the GOE model, where $\Gamma = \sqrt{D}$. No further assumptions on the nature of the system and initial state are made.

Since the eigenstates in the bulk of the spectrum are nearly random vectors, they are statistically independent from the eigenvalues. This fact is used in the derivation of the analytical expression for $\left<P_S(t)\right>$, which is explained in detail in Appendix~\ref{sec:derivation}. The equation is given by,
\be
\left<P_S(t)\right> =\frac{1-\overline{P_S}}{ (D-1)}\left[\frac{ D e^{-\Gamma^2t^2}}{4{\cal N}^2} {\cal F}(t) -b_2\left(\frac{\Gamma t}{\sqrt{2\pi}D}\right)\right] + \overline{P_S} ,
\label{eq:PSrealistic} 
\ee
where
\be
{\cal F}(t) = \left|\textrm{erf}\left(\frac{E_\textrm{max}+it\Gamma^2}{\sqrt{2}\Gamma}\right)-\textrm{erf}\left(\frac{E_\textrm{min}+it\Gamma^2}{\sqrt{2}\Gamma}\right)\right|^2,
\ee
${\cal N}$ is a normalization constant (see Appendix~\ref{sec:derivation}), $\textrm{erf}$ is the error function, $E_\textrm{max}$ is the largest eigenvalue of $H$ and $E_\textrm{min}$ is the lowest eigenvalue.

In addition to the asymptotic value $\overline{P_S}$, Eq.~(\ref{eq:PSrealistic}) contains two other terms. The one with ${\cal F}(t)$ describes the initial decay of the survival probability. At short times, the decay follows a Gaussian, $\sim e^{-\Gamma^2t^2}$, up to $t\sim 1/\Gamma$, which is the characteristic time for the depletion of the initial state. Later, when the bounds of the spectrum are reached, this term behaves like a power law $\propto t^{-2}$~\cite{Tavora2016,Tavora2017,TorresARXIV}:
\be
\frac{De^{-\Gamma^2t^2 } }{4{\cal N}^2} {\cal F}(t)  \rightarrow  \frac{D}{\Gamma^2 t^2} \quad \textrm{for }  \Gamma t\gg  1.
\label{Eq:powerlaw}
\ee
At yet longer times, the dynamics become dominated by the $b_2$ function. Its functional form is the same as that for the GOE in Eq.~(\ref{eq:b2}), because the level statistics of realistic chaotic models described by real symmetric Hamiltonian matrices are comparable to those for the GOE. 
We reiterate that up to this point, no specific model was considered.

\begin{figure}[h!]
\includegraphics[width=0.8\columnwidth]{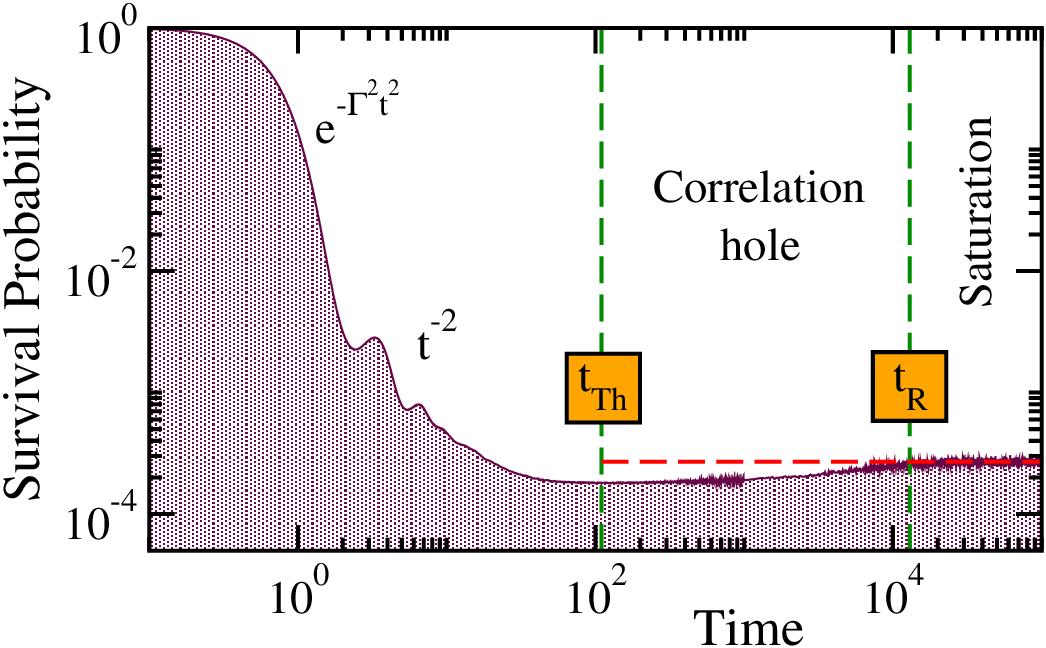}
\caption{Different stages of the evolution of the survival probability for a realistic chaotic model with local two-body interaction and initial states very delocalized in the energy eigenbasis. Same model and parameters as in Fig.~\ref{fig:SP}~(a) with $L=16$.}
\label{fig:sketch}
\end{figure}

Figure~\ref{fig:sketch} illustrates the entire evolution of the survival probability for a generic chaotic many-body model that satisfies the conditions described above. Up to $t_{\text{Th}}$, which marks the minimum of the correlation hole, the dynamics differ from what we have for the GOE matrices in Fig~\ref{fig:GOE}. Here, a Gaussian behavior and a power-law decay $\propto t^{-2}$ are observed.  Universality, in the form of the correlation hole, takes place only beyond $t_{\text{Th}}$. The dynamics saturate at $t_{\text{R}}$, after which there are only fluctuations around the infinite time average $\overline{P_S}$.

\subsection{Analytical estimation for the Thouless time and relaxation time}
\label{sec:realisticscales}

With Eq.~(\ref{eq:PSrealistic}), one can obtain analytical estimates for the time of the minimum of the hole $t_{\text{Th}}$ and for the relaxation time $t_{\text{R}}$, following the procedure of Sec.~\ref{sec:GOE}. 

\subsubsection{Thouless time}
To obtain $t_{\text{Th}}$, we expand the first term in Eq.~(\ref{eq:PSrealistic}) for long times, as done in Eq.~(\ref{Eq:powerlaw}), which gives the power-law decay $\propto t^{-2}$. And we expand the $b_2$ function to short times, which gives the linear increase in $t$,
\be
b_2\left(\frac{\Gamma t}{\sqrt{2\pi}D}\right) \rightarrow 1 - 2\frac{\Gamma t}{\sqrt{2\pi}D} \hspace{0.4 cm} \text{for} \hspace{0.4 cm} \frac{\Gamma t}{D}\ll1.
\ee
Combining the expansion in Eq.~(\ref{Eq:powerlaw}) and the expansion above in the derivative of $\langle P_S(t)\rangle$, we arrive at one of our central results,
\be
t_{\text{Th}} \propto \frac{D^{2/3}}{\Gamma}\sim \frac{e^{2cL/3}}{\sqrt{L}},
\label{eq:tThrealistic}
\ee
where we used that the Hilbert space dimension of the system is $D\propto e^{cL}$, for some constant $c>0$.
This result for $t_{\text{Th}}$ is completely different from what we have for the GOE model in Eq.~(\ref{Eq:tthGOE}). While for full random matrices, $t_{\text{Th}}^{\text{GOE}}$ is independent of system size, for realistic chaotic systems $t_{\text{Th}}$ grows exponentially with $L$. Such exponential increase of $t_{\text{Th}}$ is a general result for realistic many-body quantum systems with local interactions. Mathematically, this is caused by two combined factors: the rate of the initial Gaussian decay of $\left<P_S(t)\right>$ increases just linearly with $L$, because the Hamiltonian matrices describing real systems are sparse, and this decay is followed by a power-law behavior that lasts for longer as $L$ grows. 

In noninteracting models, the time that it takes for a particle to diffusively cross a disordered medium is called Thouless time. For realistic interacting quantum systems, we use the same terminology to denote the time for $\left<P_S(t)\right>$ to reach the minimum of the correlation hole. The region of the correlation hole is exclusively present in finite quantum systems with a discrete spectrum and correlated eigenvalues. It takes the time $t_\textrm{Th}$ for the dynamics to resolve the discreteness of the spectrum and detect spectral correlations. After $t_\textrm{Th}$, the dynamics consist purely of dephasing processes, and are fully quantum in nature.
 
Physically, we interpret the Thouless time in interacting systems as the time for the initial many-body state to spread over an exponentially large many-body Hilbert space via local interactions, which takes an exponentially long time. The initially localized state, given by one eigenstate of the unperturbed Hamiltonian $H_0$, needs time $t_\textrm{Th}$ to acquire weight over all many-body states of $H_0$ in the microcanonical energy shell. This contrasts with the GOE model, where the initial state is directly coupled with all eigenstates of $H_0$, so the time to reach the minimum of the correlation hole does not depend on system size.

\begin{figure}[h!]
\includegraphics[width=0.8\columnwidth]{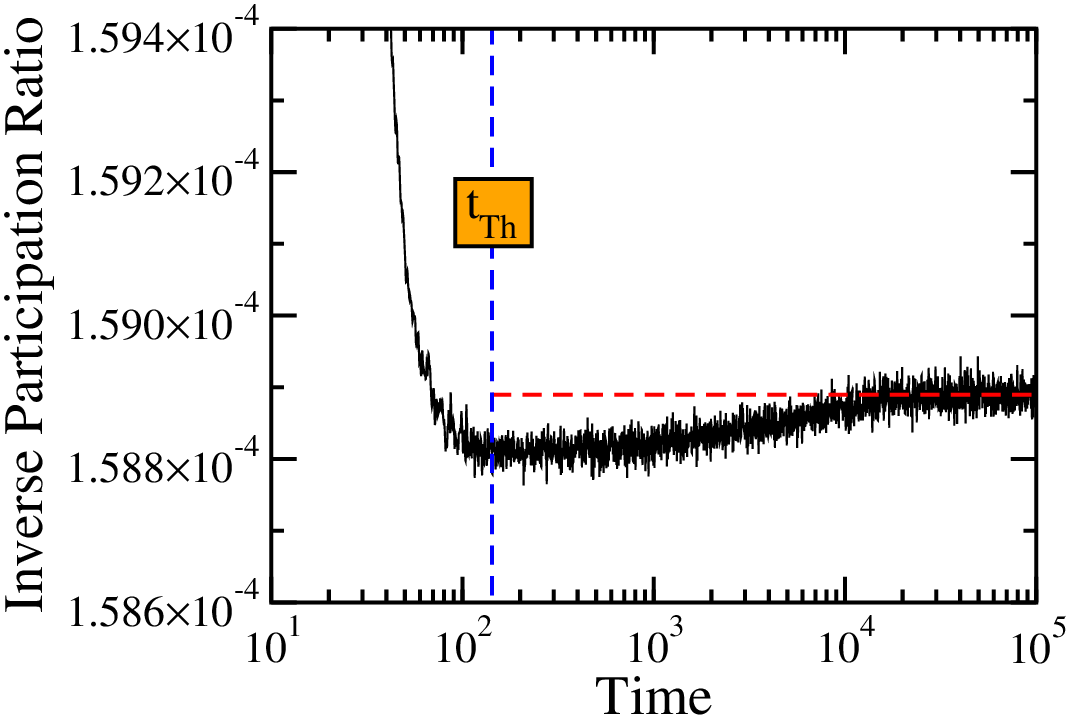}
\caption{Spread in time of an initially localized state through the many-body Hilbert space. The spread is  quantified by the inverse participation ratio. In the figure, $\left<\text{IPR} (t) \right>$ is multiplied by the dimension $D$ of the Hilbert space. Same realistic chaotic model with local two-body interaction used in Fig.~\ref{fig:sketch} and in Fig.~\ref{fig:SP}~(a) with $L=16$. The vertical dashed line marks the Thouless time and the horizontal dashed line indicates the saturation value.}
\label{fig:IPR}
\end{figure}

To describe the spread of the initial state in the many-body space of a realistic system, we compute the evolution of the inverse participation ratio,
\begin{equation}
\left<\text{IPR} (t) \right> =\sum_{n}\left| \langle \phi_n | e^{-iHt} |\Psi(0)\rangle \right|^4,
\end{equation}
which quantifies the inverse of the number of unperturbed many-body states that contribute to the dynamics. When $\left<\text{IPR} (t) \right>$ reaches its minimal value, the spreading of the initial state in the Hilbert space is maximal. This is illustrated in Fig.~\ref{fig:IPR} for the same generic chaotic many-body model considered in Fig.~\ref{fig:sketch}. Figure~\ref{fig:IPR} confirms that the minimum of $\left<\text{IPR} (t) \right> $, just as the minimum of  $\left<P_S(t)\right>$, happens at $t_{\text{Th}}$. 

\subsubsection{Relaxation time}
We now examine the relaxation time $t_R$. For long times, 
\be
b_2\left(\frac{\Gamma t}{\sqrt{2\pi}D}\right) \rightarrow\frac{\pi D^2}{6\Gamma^2t^2} 
\hspace{0.4 cm} \text{for} \hspace{0.4 cm} \frac{\Gamma t}{D}\gg1.
\ee
Since the term above is proportional to $D^2$, while Eq.~(\ref{Eq:powerlaw}) is proportional to $D$, we can discard the latter for large $D$. Following the same procedure as in  Sec.~\ref{sec:tR}, one finds that
\be
t_\text{R}\propto\frac{D}{\Gamma\sqrt{\delta}}\sim\frac{e^{cL}}{\sqrt{L\delta}}.
\label{eq:tRrealistic}
\ee
Since for realistic chaotic systems and for the GOE model, the dynamics at long times are dominated by the same function $b_2$, we obtain again that $t_\text{R}$ is inversely proportional to the mean level spacing. This result demonstrates analytically that the time beyond which the observable simply fluctuates around the infinite-time average is the Heisenberg time. 

By comparing Eqs.~(\ref{eq:tThrealistic}) and~(\ref{eq:tRrealistic}), one sees that as the system size $L$ grows, the Thouless and relaxation times move exponentially far apart from each other and the correlation hole gets elongated.

\subsection{Numerical results for different realistic chaotic models}
\label{sec:comp}

\begin{figure}[h!]
\includegraphics[width=1\columnwidth]{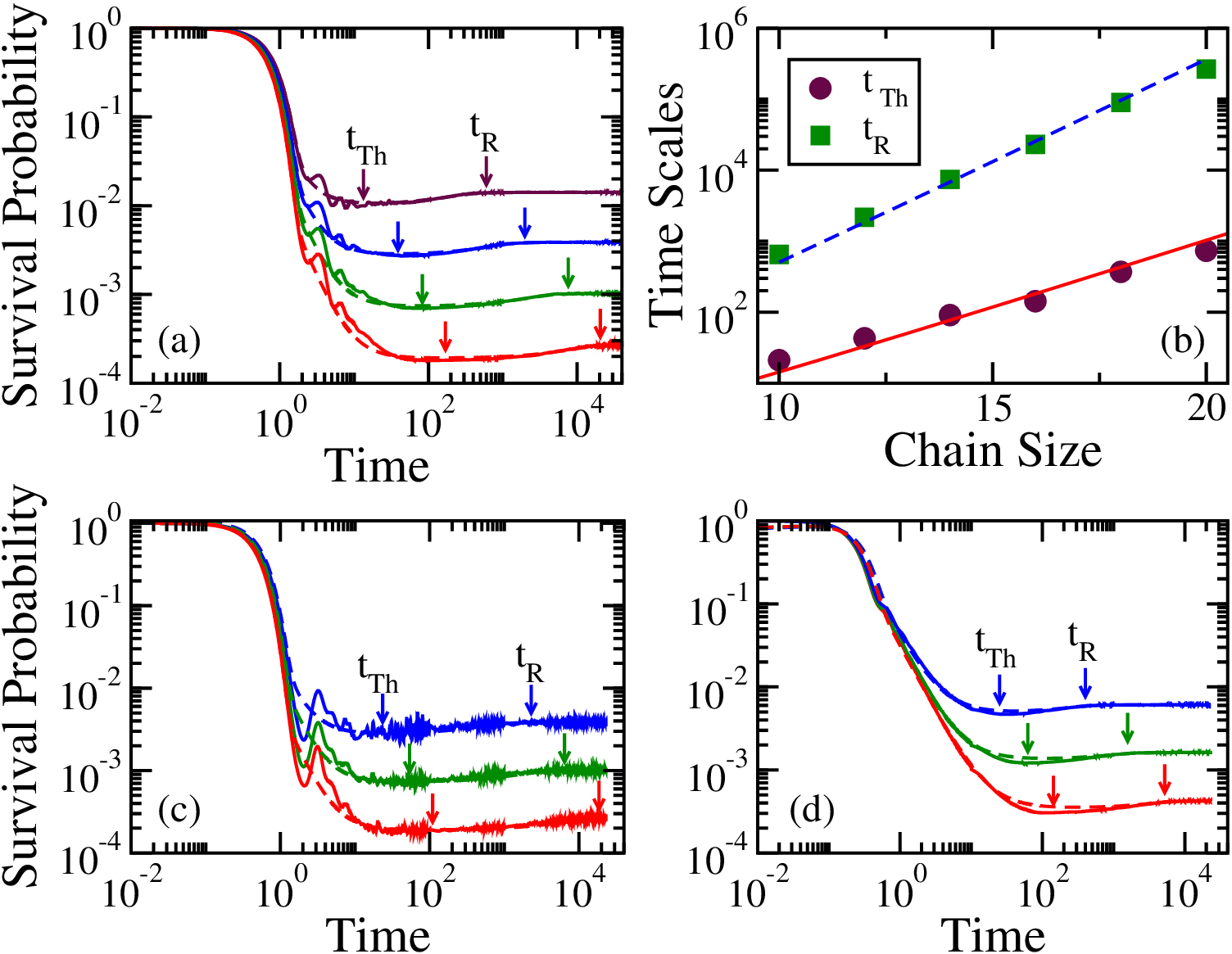}
\caption{Survival probability for realistic chaotic models. In (a), (c) and (d) we compare numerical data (full lines) with the analytical  Eq.~(\ref{eq:PSrealistic}) (dashed lines). (a) Disordered spin-1/2 model from Eq.~(\ref{eq:HNN}) with disorder strength $h=0.5$ and system sizes $L=10,12,14,16$ from top to bottom. (c) Clean spin-1/2 model with next-to-nearest-neighbors couplings from Eq.~(\ref{Eq:clean}) and system sizes $L=12,14,16$ from top to bottom. (d) Sparse banded random matrix model with matrices sizes $D=924, 3\,432, 12\,870$. In (b) we compare the values of $t_\textrm{Th}$ (circles) and $t_R$ (squares) extracted numerically for the disordered spin-1/2 model with the analytical Eqs.~(\ref{eq:tThrealistic}) and~(\ref{eq:tRrealistic}) (full and dashed lines, respectively), finding excellent agreement.}
\label{fig:SP}
\end{figure}

In Fig.~\ref{fig:SP}, we compare our analytical Eq.~(\ref{eq:PSrealistic}) for the survival probability with numerical data for three different realistic chaotic models. We use the lower bound $E_\textrm{min}$ as a single fitting parameter.  

In Fig.~\ref{fig:SP}~(a), we plot the data for a disordered  spin-1/2 chain with nearest-neighbor couplings only. The total Hamiltonian $H^\textrm{ds}$ has two terms,
\ba
H^\textrm{ds} &=& H_0^\textrm{ds} + V^\textrm{ds},
\label{eq:HNN}
\\
H_0^\textrm{ds}&=& J \sum_{k=1}^L (h_k  S_k^z +S^z_k S^z_{k+1}), \nonumber\\
V^\textrm{ds}&=&J\sum_{k=1}^L (S^x_k S^x_{k+1}+S^y_k S^y_{k+1}). \nonumber
\ea
Above, $S_k^{x,y,z}$ are the spin operators on site $k$, $L$ is the size of the chain, and the amplitudes $h_k$ are uniform random numbers in $[-h,h]$, $h$ being the disorder strength. We set $J=1$ and periodic boundary conditions are assumed. This system can be mapped into models of hardcore bosons and spinless fermions and has been studied experimentally in the context of many-body localization~\cite{Schreiber2015}.

The Hamiltonian $H^\textrm{ds}$ conserves the total magnetization ${\cal S}^z=\sum_kS_k^z$. We work with the largest subspace ${\cal S}^z=0$,  where the dimension of the Hilbert space is $D=L!/(L/2)!^2 \sim e^{L \ln 2}$, so $c=\ln2$ in Eq.~(\ref{eq:tThrealistic}) and in Eq.~(\ref{eq:tRrealistic}). We take the disorder strength $h=0.5$, where the model is maximally chaotic~\cite{Torres2017}. To compute $\left<P_S(t)\right>$, an average over initial states with energies close to the middle of the spectrum and over disorder realizations is performed. 

As clearly seen in Fig.~\ref{fig:SP}~(a), the analytical prediction from Eq.~(\ref{eq:PSrealistic}) describes accurately the numerical curve for $\left<P_S(t)\right>$ for more than six orders of magnitude in time, covering the entire evolution, from $t \sim 1/\Gamma$ to $t\sim t_R$. The figure shows that both $t_\textrm{Th}$ and $t_\text{R}$ grow with the system size. A more quantitative analysis is provided in Fig.~\ref{fig:SP}~(b), where we plot $t_\textrm{Th}$ and $t_\text{R}$ as a function of  $L$ and compare them with our analytical estimates in Eq.~(\ref{eq:tThrealistic}) and Eq.~(\ref{eq:tRrealistic}). The agreement is excellent. The exponential growth of both $t_\textrm{Th}$ and $t_R$ is clearly visible, as well as the growth of the difference between them, which indicates the stretch of the correlation hole with $L$.

To show that Eq.~(\ref{eq:PSrealistic}) is indeed general, we test it for two other models. In Fig.~\ref{fig:SP}~(c), we plot the survival probability for a clean spin-1/2 model with next-to-nearest-neighbors couplings. Its Hamiltonian reads
\ba
H^\textrm{cl}&=&H_0^\textrm{cl} + V^\textrm{cl},
\label{Eq:clean} \\
H_0^\textrm{cl}&=&J \Delta \sum_{k=1}^{L}\left(S_k^zS_{k+1}^z+\lambda S_k^zS_{k+2}^z\right), \nonumber \\
V^\textrm{cl}&=&J\sum_{k=1}^{L}\left[S_k^xS_{k+1}^x+S_k^yS_{k+1}^y+\lambda\left(S_k^xS_{k+2}^x+S_k^yS_{k+2}^y\right)\right].\nonumber
\ea
We choose open boundary conditions, $J=1$, anisotropy parameter $\Delta=0.48$, the strength of the next-to-nearest-neighbors coupling $\lambda=1$, and ${\cal S}^z=0$, so that again $D=L!/(L/2)!^2$.  Despite the absence of random elements, this model is strongly chaotic as well~\cite{Schiulaz2018}. The average is now performed over initial states only, which explains why the numerical data in Fig.~\ref{fig:SP}~(c) show larger fluctuations than for the disordered spin model in Fig.~\ref{fig:SP}~(a). The analytical curves for different system sizes capture the numerical behavior of $\left<P_S(t)\right>$ extremely well.

As a third example, in Fig.~\ref{fig:SP}~(d), we plot the data for a sparse banded random matrix model. This model has the same nonzero entries as the Hamiltonian in Eq.~(\ref{Eq:clean}), but they are drawn independently from a Gaussian distribution with mean value $0$ and variance $J^2$. An average over  initial states with energies at the middle of the spectrum and over several realizations of the Hamiltonian is performed. This model is not related to any specific physical system. Once again, the numerical evolution of $\left<P_S(t)\right>$ follows very well the analytical expression.

\section{Transition from chaos to localization} 
\label{sec:numerics}

In the previous section, we considered only systems in the strongly chaotic regime. It is now natural to ask how the results change for systems away from this regime. In this section, we analyze this question for the disordered spin-1/2 model of Eq.~(\ref{eq:HNN}). At a critical value $h_c > 2.25$, this system transitions to a many-body localized phase, where the eigenvalues are uncorrelated. We consider disorder strengths $0.5 \leq h \leq 2.25$, where the energy levels have some degree of correlation. We find that as $h$ is increased above 0.5, the Thouless time progressively approaches the relaxation time until their values coincide and the correlation hole disappears. 

\subsection{Growth of the Thouless time with disorder}

\begin{figure}[h!]
\includegraphics[width=1\columnwidth]{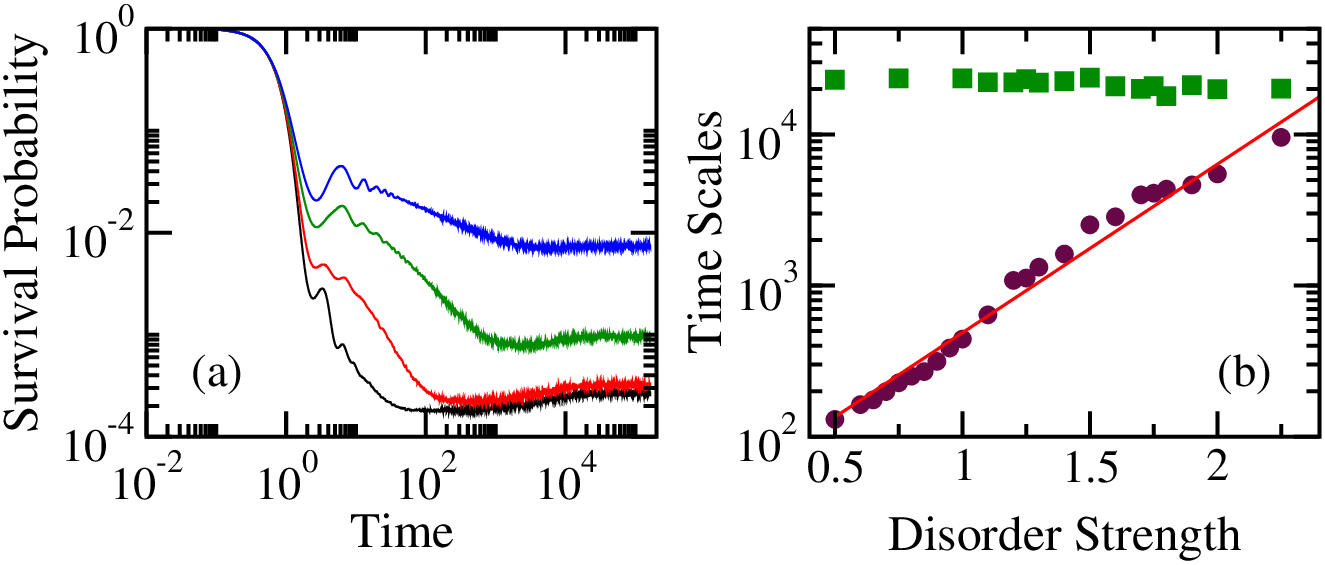}
\caption{Survival probability (a) and long timescales (b) for the disordered spin-1/2 model (\ref{eq:HNN}) with different disorder strengths. (a) $\left<P_S(t)\right>$ for $h=0.5,1.0,1.5,2.0$, from bottom to top, and system size $L=16$. (b) Thouless time (circles) and relaxation time (squares) as a function of disorder strength. The solid line shows the fit $t_{\text{Th}}\sim 37 e^{2.6h}$.}
\label{fig:disorder}
\end{figure}

In Fig.~\ref{fig:disorder}~(a), we plot the survival probability for different disorder strengths, increasing from bottom to top, at system size $L=16$. The consequence of the presence of disorder is  different at different timescales. For short times, where the Gaussian decay holds, the disorder has no effect on the dynamics, because $\Gamma$ depends only on the off-diagonal entries of the Hamiltonian, which are independent of $h$. For $\Gamma t \lesssim1$, all curves fall on top of each other.  At later times, in the region of the power-law decay, the power-law exponent decreases as a function of $h$, as explained in Ref.~\cite{Torres2015,Torres2017}. At even later times, the $b_2$ function is also affected by disorder: the correlation hole gets delayed and $t_\textrm{Th}$ grows as $h$ increases. Finally, while the saturation value $\overline{P_S}$ naturally increases as the disorder strength increases, since the initial states become less spread out in the energy eigenbasis, the time $t_R$ at which such value is reached does not change. This is because $t_R$ is inversely proportional to the mean level spacing, which does not strongly depend on disorder for $0.5 \leq h \leq 2.25$.

The dependence of the long timescales on the disorder strength can be seen more quantitatively in Fig.~\ref{fig:disorder}~(b), where we plot $t_{\text{Th}}$ and $t_R$ as a function of $h$. The Thouless time grows exponentially with $h$, indicating that the spread of the initial state in the many-body space becomes much slower. $t_{\text{Th}}$ eventually reaches $t_{\text{R}}$ for $h>2.25$, when the system localizes and the correlation hole ceases to exist. We do not show data for this region, because for $h\gtrsim2.25$, the hole becomes tiny and it becomes challenging to distinguish numerically the Thouless time from the relaxation time. 

We notice that, in noninteracting disordered systems, the ratio $t_{\text{R}}/t_{\text{Th}}$ is called Thouless dimensionless conductance. It is large in the metallic phase and it approaches 1 as the system approaches the localized phase. For the interacting disordered spin model from Eq.~(\ref{eq:HNN}) in the chaotic regime, our results show that  $t_{\text{R}}/t_{\text{Th}} \propto e^{L (\ln2)/3}$. As the disorder strength grows and the system leaves the chaotic region toward many-body localization, the gap between the two timescales decreases exponentially with $h$ and $t_{\text{R}}/t_{\text{Th}}\rightarrow 1$. This ratio is thus an additional tool for the studies of localization in interacting systems.

\subsection{Relation between the Thouless time and the Thouless energy}

In noninteracting disordered systems, the Thouless time was originally defined as the diffusion time of a particle through the sample. It is inversely proportional to the Thouless energy, $E_{\text{Th}}$, which is determined by the diffusion constant and the system size~ \cite{Thouless1974,Guhr1998}. Later, it was shown that, within the energy scale defined by $E_{\text{Th}}$, the level statistics of these systems follow those from random matrices~\cite{Altshuler1986,*Altshuler1988}. The analysis of level statistics can then be used as an alternative way to identify the Thouless energy. 

Here, we investigate how this picture can be extended to interacting systems. For our definition of the Thouless time, namely the time to reach the minimum of the correlation hole, we indeed recover that $t_{\text{Th}} \propto 1/E_{\text{Th}}$. But before showing these results, let us explain how $E_{\text{Th}}$ is obtained from the spectral correlations of chaotic models.

The energy levels of  chaotic systems are strongly correlated. Long-range correlations can be quantified by computing the level number variance $\Sigma^2(\ell)$. This is done as follows. One first has to unfold the spectrum, in order to set the smooth part of the density of states to a constant~\cite{Guhr1998}. Then, one partitions the spectrum into intervals of length $\ell$, counts the number of levels inside each interval, and computes the variance of the resulting distribution.
For GOE random matrices, strong correlations between the eigenvalues manifest as a logarithmic growth for the level number variance,
$
\Sigma^2(\ell)=\frac{2}{\pi^2}\left[\log(2\pi\ell)+\gamma_e+1-\frac{\pi^2}{8}\right],
$
where $\gamma_e=0.5772\cdots$ is the Euler-Mascheroni constant. 

For chaotic noninteracting disordered models, it was found in~\cite{Altshuler1986} that $\Sigma^2(\ell)$ grows logarithmically with the energy interval $\ell$ for $\ell<E_\textrm{Th}$, where $E_\textrm{Th}$ is the Thouless energy. For level separations larger than the Thouless energy, $\Sigma^2(\ell)$ deviates from this behavior. This notion of the Thouless energy was extended to the interacting disordered model of Eq.~(\ref{eq:HNN}) in Ref.~\cite{Bertrand2016}. There, it was shown that the Thouless energy becomes smaller as the disorder strength increases and the system approaches a many-body localized phase.

\begin{figure}[h!]
\includegraphics[width=1\columnwidth]{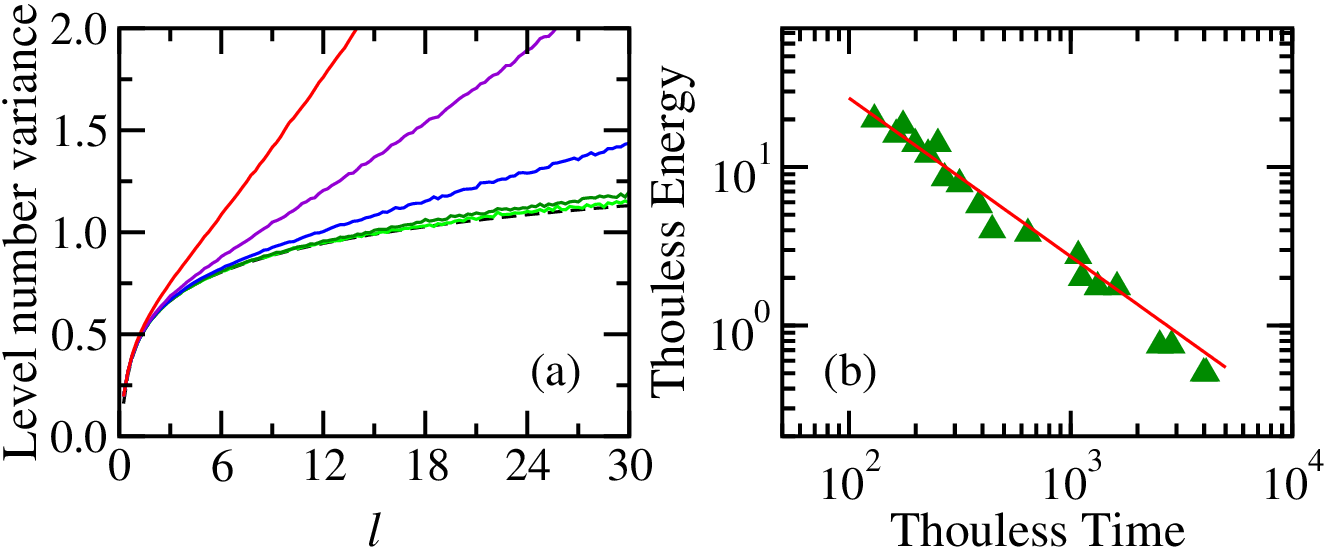}
\caption{Level number variance (a) and relation between the Thouless energy and the Thouless time (b) for the disordered spin model from Eq.~(\ref{eq:HNN}) with different disorder strengths.
(a) The analytical GOE curve (dashed line) for $\Sigma^2(\ell)$ is compared with numerical results (solid lines) for $h=0.5, 0.75, 1, 1.25, 1.5$ from bottom to top. (b) The numerical data (triangles) are fitted with $E_{\text{Th}}=2724/t_{\text{Th}}$ (solid line), showing that the Thouless energy and the Thouless time are inversely proportional to each other. Both panels: $L=16$.}
\label{fig:disorderB}
\end{figure}

In Fig.~\ref{fig:disorderB}~(a), we compare the data for $\Sigma^2(\ell)$ for various disorder strengths with the analytical GOE curve (dashed line). The Thouless energy is extracted as the point at which $\Sigma^2(\ell)$ deviates from the logarithmic behavior. In Fig.~\ref{fig:disorderB}~(b), we then analyze the relationship between $E_{\text{Th}}$ and $t_{\text{Th}}$ for various values of $h$ and confirm that $E_{\text{Th}} \propto 1/t_{\text{Th}}$ for our interacting model. This further justifies referring to the time to reach the minimum of the correlation hole as the Thouless time.

\section{Spin Autocorrelation Function}
\label{sec:I}

The survival probability and the inverse participation ratio shown in Fig.~\ref{fig:IPR} are non-local quantities. In this section, we investigate the long timescales for the spin autocorrelation function, which is a local observable in real space. It is given by
\be
I(t)=\frac{4}{L}\sum_{i=1}^L\left<\Psi_0\right|S_i^ze^{iHt}S_i^ze^{-iHt}\left|\Psi_0\right>.
\ee
This quantity measures how close the spin configuration at time $t$ is to the initial one. It is analogous to the density imbalance measured in experiments with cold atoms~\cite{Schreiber2015}. 

\begin{figure}[t!]
\includegraphics[width=1\columnwidth]{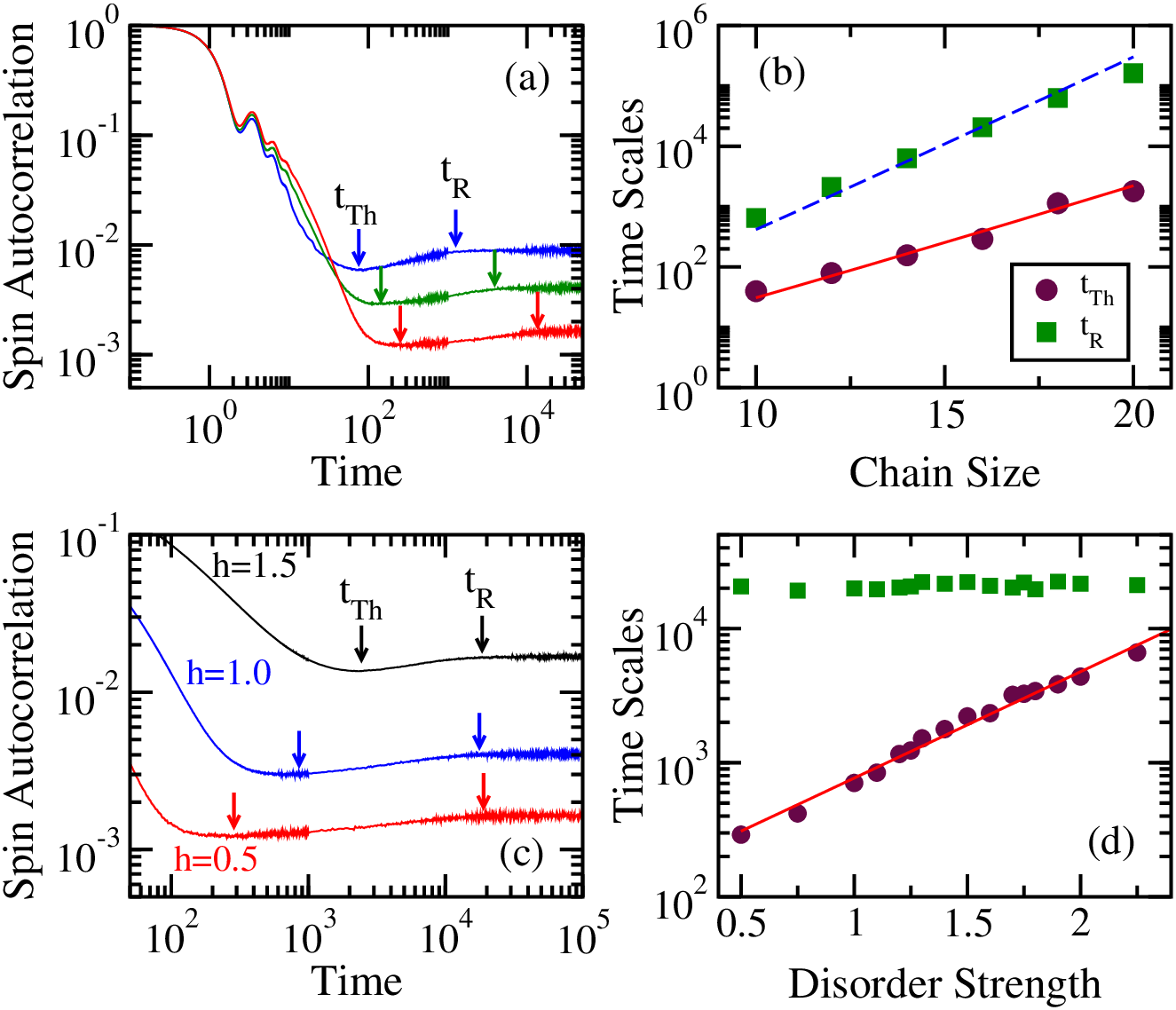}
\caption{Spin autocorrelation function for the spin model. $\left<I(t)\right>$ in (a) and (c). Thouless and relaxation times as a function of system size (b) and of disorder strength (d). Circles are for $t_{\text{Th}}$, squares for $t_{\text{R}}$. (b) Solid line is for Eq.~(\ref{eq:tThrealistic}) and dashed line for Eq.~(\ref{eq:tRrealistic}). (d) Solid line is for the fit $124 e^{1.8 h}$.  (a) $L=12,14,16$ from top to bottom.  (a) and (b) $h=0.5$. (c) and (d) $L=16$.}
\label{fig:Imb}
\end{figure}

At long times, the behavior of $\left<I(t)\right>$ is remarkably similar to $\left<P_S(t)\right>$, as seen in Fig.~\ref{fig:Imb}~(a). There, a correlation hole is also clearly visible. In Fig.~\ref{fig:Imb}~(b), we plot the numerical values for $t_{\text{Th}}$ and $t_R$ {\em vs} $L$ for the spin autocorrelation function. It shows again that the time to reach the minimum of the correlation hole increases exponentially with system size. The same estimate found for the survival probability in Eq.~(\ref{eq:tThrealistic}) matches very well the numerical results for $\left<I(t)\right>$. The time to later relax to the infinite-time average follows again Eq.~(\ref{eq:tRrealistic}), that is, it is given by the inverse of the mean level spacing. This shows that the long timescales that we unveiled for global quantities can manifest themselves for local experimental quantities as well.

Evidently, the short-time evolution of the spin autocorrelation function is different from the survival probability, as one can see by comparing Figs.~\ref{fig:SP}~(a) and~\ref{fig:Imb}~(a).  Up to $t_\text{Th}$ the dynamics depend on the initial state, model, and observable. Beyond the minimum of the correlation hole, as mentioned at different occasions in this work, the dynamics become universal and governed by spectral properties. It may happen, however,  that the amplitude of the dynamical effects caused by correlated eigenvalues is not large, as seen for $\left<\text{IPR} (t) \right>$ in Fig.~\ref{fig:IPR}. Open questions include why this happens and which observables have pronounced correlation holes, as the survival probability and the spin autocorrelation function. Another interesting question is whether for the observables with visible correlation holes, the time to reach the minimum value always follows Eq.~(\ref{eq:tThrealistic}). This is indeed what our results indicate, where the particular features of the short-time evolution of $\left<I(t)\right>$ conspire to achieve the same $L$-dependence for $t_{\text{Th}}$ as for $\left<P_S(t)\right>$.

The analogy between the spin autocorrelation function and the survival probability extends also to the transition region between chaos and localization. Just as for the survival probability, the minimum of the correlation hole for $\left<I(t)\right>$ gets postponed to later times as $h$ increases, as illustrated in Fig.~\ref{fig:Imb}~(c). This time grows exponentially with $h$, as shown in Fig.~\ref{fig:Imb}~(d), until $t_{\text{Th}} \sim t_{\text{R}}$. Therefore, the analysis of how the ratio $t_{\text{R}}/t_{\text{Th}}$ approaches 1 may be used to detect the transition to localization also when local observables are considered.

The fact that the time to achieve complete relaxation increases exponentially with system size, be the observable global or local, is of consequence to theoretical and experimental studies of relaxation and thermalization. Needless to say, reaching $t_\text{Th}$ or $t_\text{R}$ experimentally is challenging. However, coherence times are being pushed to ever longer values. In particular, the Thouless time for systems with $L\leq 18$ might soon be within reach.


\section{Conclusion}
\label{sec:conclusion}

This work promotes the use of dynamical manifestations of spectral properties, which emerge when the time evolution resolves the discreteness of the spectrum, as a means to identify the long timescales involved in the relaxation process of interacting many-body quantum systems. In doing so, we find that there is not only one, but two very long timescales: the Thouless time, $t_\text{Th}$, and the relaxation time, $t_\text{R}$.

We derive analytical estimates for $t_\text{Th}$ and $t_\text{R}$ for realistic interacting systems in the chaotic regime. They match extremely well our numerical results for a global quantity and an experimental local observable. These are the survival probability and the spin autocorrelation function, respectively. 

We provide a physical interpretation for the Thouless time in interacting systems. When interactions are present, the dynamics cannot be completely captured in terms of real space processes, but require instead the analysis of the evolution in the  many-body Hilbert space. Using the inverse participation ratio, we showed that $t_\text{Th}$ corresponds to the time for a many-body initial state to get completely spread out, via local interactions, in the many-body Hilbert space. Since this space is exponentially large in the system size $L$, the Thouless time grows exponentially with $L$. This is to be contrasted with our results for the GOE model, where the matrices are fully connected and $t_{\text{Th}}^{\text{GOE}}$ is therefore independent of the matrix size.

Our derivations demonstrate that the relaxation time coincides with the Heisenberg time, being thus the largest timescale of the system dynamics. The analytical estimate for $t_\text{R}$ is the same for realistic systems and for the GOE model, since the dynamics beyond $t_{\text{Th}}$ become universal.

In noninteracting disordered systems, the ratio between the Heisenberg time and the Thouless time is the Thouless dimensionless conductance, which  goes to 1 as the system approaches the localized phase. This prompts us to use the disordered interacting spin model to analyze $t_{\text{R}}/t_{\text{Th}}$, finding that the ratio approaches 1 exponentially fast with the disorder strength.
We verify that the parallel between interacting and noninteracting disordered systems extends also to the relationship between the Thouless time and the Thouless energy. We find that $t_{\text{Th}} \propto 1/E_{\text{Th}}$, which gives further support to our definition of the Thouless time.

Definitions of the Thouless time based on transport properties~\cite{Serbyn2017,Varma2017,DymarskyARXIV,DymarskyARXIVThouless} lead to a power-law scaling of $t_\textrm{Th}$ with system size. This result does not agree with our definition, which is based on the dynamical manifestations of spectral correlations. While these two approaches coincide for noninteracting systems, they are not equivalent for interacting many-body systems. Understanding this discrepancy is a critical point for future works on nonequilibrium many-body quantum dynamics and related subjects, such as many-body localization, many-body quantum chaos, and thermalization.

\begin{acknowledgments}
M.S. and L.F.S. are supported by the NSF Grant No.~DMR-1603418. E.J.T.-H. acknowledges funding from VIEP-BUAP (Grant Nos.~MEBJ-EXC19-G and LUAG-EXC19-G), Mexico. He is also grateful to LNS-BUAP for allowing use of their supercomputing facility.  We are very thankful to Francisco P\'erez-Bernal for allowing us to use the supercomputer at the University of Huelva in Spain and for providing technical assistance. 
\end{acknowledgments}

\appendix

\section{Derivation of the expression for the survival probability for realistic many-body quantum systems}
\label{sec:derivation}

Here, we show the steps to obtain Eq.~(\ref{eq:PSrealistic}), which describes the entire evolution of the averaged survival probability. We reiterate that Eq.~(\ref{eq:PSrealistic}) is general and valid for realistic many-body quantum systems on a finite lattice, which are strongly chaotic, present only local two-body interactions, and are perturbed very far from equilibrium ({\em i.e.} beyond the Fermi golden rule regime). The initial states correspond to site-basis vectors (computational basis vectors) with energies away from the edges of the spectrum, so that they are highly delocalized in the energy eigenbasis. 

The equation for the survival probability can be written in the following forms,
\ba
P_S(t) &=& \left| \langle \Psi(0) | e^{-iHt} | \Psi(0) \rangle \right|^2 = 
\left| \sum_{\alpha} \left| C_\alpha^{(0)} \right|^2 e^{-i E_\alpha t} \right|^2 \nonumber \\
&=&\sum_{\alpha \neq \beta}\left|C_\alpha^{(0)}\right|^2\left|C_\beta^{(0)}\right|^2e^{-i(E_\alpha-E_\beta)t}  + \sum_{\alpha} \left| C_\alpha^{(0)} \right|^4 \nonumber \\
&=& \int G(E) e^{-i E t} dE,
\label{Eq:Fourier}
\ea
where $C_\alpha^{(0)}=\left<\alpha|\Psi(0)\right>$ and the integrand $G(E)$ is
\ba
G(E) &=&\sum_{\alpha\neq\beta}\left|C_\alpha^{(0)}\right|^2\left|C_\beta^{(0)}\right|^2 \delta (E - E_{\alpha} + E_{\beta} ) \nonumber \\
&+&
 \sum_{\alpha}\left|C_\alpha^{(0)}\right|^4 \delta (E).
\label{Eq:GE}
\ea
This function is similar to the spectral autocorrelation function, $\sum_{\alpha,\beta} \delta (E - E_{\alpha} + E_{\beta} )$, the difference being the weights $|C_\alpha^{(0)}|^2$.

To obtain the averaged survival probability, 
\be
\langle P_S(t) \rangle
= \int \langle G(E) \rangle e^{-i E t} dE,
\ee
we take into account the asymptotic value, 
\begin{equation}
\overline{P_S} =\left<\sum_{\alpha}\left|C_\alpha^{(0)}\right|^4\right>,
\end{equation}
and need to compute
\be
\left\langle G(E) \right\rangle_{\alpha\neq\beta} = \left\langle \sum_{\alpha\neq\beta}\left|C_\alpha^{(0)}\right|^2\left|C_\beta^{(0)}\right|^2 \delta (E - E_{\alpha} + E_{\beta} ) \right\rangle.
\ee


\subsection{Factorization of eigenvalues and eigenvectors}
\label{Sec:fact}

In full random matrices, where the eigenstates are random vectors and the coefficients are then uncorrelated random numbers, the eigenvalues and eigenstates are statistically independent, which allows for the factorization \cite{Alhassid1992,SantosTorres2017AIP,TorresARXIV},
\be
\left<G(E)\right>_{\alpha\neq\beta}=\sum_{\alpha\neq\beta}\left<\left|C_\alpha^{(0)}\right|^2\left|C_\beta^{(0)}\right|^2\right>\left<\delta(E-E_\alpha+E_\beta)\right>.
\label{Eq:GEapp}
\ee

For realistic chaotic many-body quantum systems, it is reasonable to expect a similar (but not identical) scenario, provided they are perturbed very far from equilibrium and the initial state has energy close to the middle of the spectrum, {\em i.e.} $E_0 \sim 0$, as indeed considered in our work. In the bulk of the spectrum, the eigenstates are chaotic~\cite{Torres2014PRA,Torres2014NJP,Torres2014PRE,Torres2014PRAb}, while states close to the edges of the spectrum are not. By chaotic states, we mean states for which the coefficients are (nearly) uncorrelated and fill the entire energy shell~\cite{Santos2012PRL,*Santos2012PRE,Borgonovi2016}. In the limit of very strong perturbation, beyond the Fermi golden rule regime, initial states with $E_0 \sim 0$ fall within the chaotic region of the spectrum, being themselves chaotic states, so the majority of their components $|C_\alpha^{(0)}|^2$ are nearly uncorrelated.

\begin{figure}[ht]
\includegraphics[width=0.9\columnwidth]{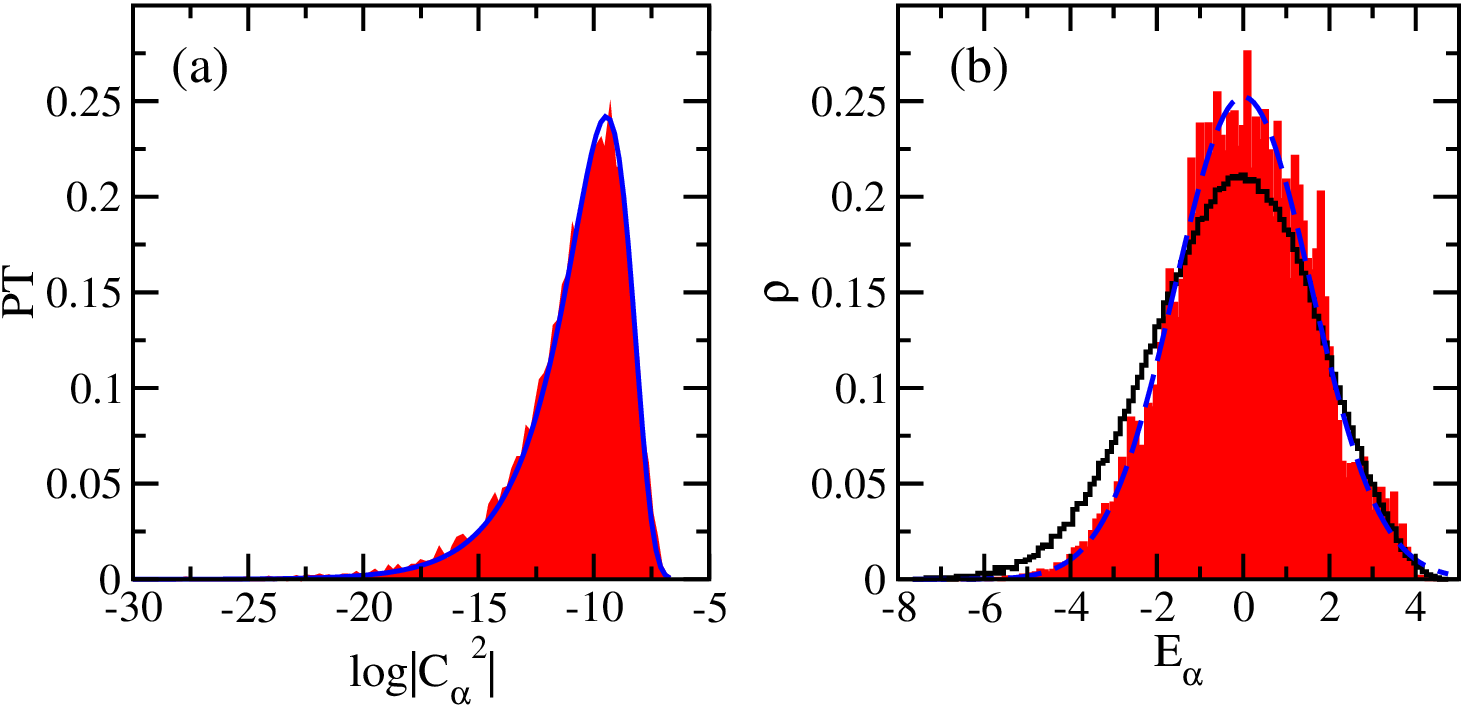}
\caption{Shaded areas: (a) Distribution of the coefficients $|C^{(0)}_\alpha|^2$ for a single initial state with energy in the middle of the spectrum and (b) energy distribution of this initial state (LDOS). (a) The solid line is the Porter-Thomas distribution given in Eq.~(\ref{eq:PT}). (b) The solid line is the density of states and the dashed line is the Gaussian fit for the LDOS. Disordered spin-1/2 model described in Eq.~(\ref{eq:HNN}) with size $L=16$ and disorder strength $h=0.5$.}
\label{fig:C}
\end{figure}

To further support the assumption of the chaoticity of the initial state, we study in Fig.~\ref{fig:C}~(a) the distribution of its components $|C_\alpha^{(0)}|^2$. In random matrix theory, the components of chaotic states are known to follow the Porter-Thomas distribution~\cite{Brody1981},
\be
PT \left(\left|C_\alpha^{(0)}\right|^2\right) = \left( \frac{D}{2 \pi \left|C_\alpha^{(0)}\right|^2 } \right)^{1/2}  \exp \left( - \frac{D}{2} \left|C_\alpha^{(0)}\right|^2 \right) .
\label{eq:PT}
\ee
As seen in Fig.~\ref{fig:C}~(a), this is indeed the distribution obeyed by $|C_\alpha^{(0)}|^2$ for the chaotic disordered spin-1/2 model from Eq.~(\ref{eq:HNN}).  Notice that it holds even though we consider in the figure a single initial state and a single disorder realization.

The explanations above justify proceeding with the factorization in Eq.~(\ref{Eq:GEapp}), although corrections do exist. For instance, while both the energy distribution of the initial state (LDOS), 
\[
\rho_{0}(E) = \sum_{\alpha} |C_\alpha^{(0)}|^2 \delta(E-E_\alpha )
\]
 and the density of states, 
\[
 R_1(E)=\sum_{\alpha}  \delta(E-E_\alpha ) 
\]
have a  Gaussian shape, as expected for many-body quantum systems with two-body couplings~\cite{French1970,Brody1981}, the LDOS is narrower than the density of states. This is clearly seen in Fig.~\ref{fig:C}~(b). However, as our numerical results in Sec.~\ref{sec:comp} show, these corrections do not affect the general features of the initial decay of the survival probability, only details that are not relevant for our estimates of the times cales obtained in Sec.~\ref{sec:realisticscales}.

Since the average over the components of the initial state is 
\be
\left<\sum_{\alpha\neq\beta}\left|C_\alpha^{(0)}\right|^2\left|C_\beta^{(0)}\right|^2\right>=\left<1-\sum_{\alpha}\left|C_\alpha^{(0)}\right|^4\right>=1-\overline{P_S},
\label{eq:const}
\ee
we are left with
\be
\langle P_S(t) \rangle
= \left( 1 - \overline{P_S} \right)
\int  \left<\delta(E-E_\alpha+E_\beta)\right>  e^{-i E t} dE + \overline{P_S} .
\label{Eq:FourierApp}
\ee
To compute the integral above, we use the fact that the average over the spacing distributions can be written in terms of the two-point spectral correlation function $R_2(E_\alpha,E_\beta)$, as~\cite{MehtaBook,Alhassid1992}
\ba
\label{eq:SM_delta}
\left<\delta(E-E_\alpha+E_\beta)\right>&=&\frac{(D-2)!}{D!}\int dE_\alpha dE_\beta \\
&\times&\delta(E-E_\alpha+E_\beta)R_2(E_\alpha,E_\beta).\nonumber
\ea
The function $R_2(E_\alpha,E_\beta)$ can be decomposed into the density of states $R_1(E_\alpha)$ and the two-level cluster function $T_2(E_\alpha,E_\beta)$, so that
\be
R_2(E_\alpha,E_\beta)=R_1(E_\alpha)R_1(E_\beta)-T_2(E_\alpha,E_\beta).
\label{Eq:R2_SM}
\ee

\subsection{Gaussian density of states}

Plugging the first term of Eq.~(\ref{Eq:R2_SM}) into the Fourier transform in Eq.~(\ref{Eq:FourierApp}) gives
\ba
&&\frac{(D-2)!}{D!}   \times \nonumber \\
&&\int e^{-iEt}  \delta(E-E_\alpha+E_\beta)R_1(E_\alpha)R_1(E_\beta) \, dE \, dE_\alpha \, dE_\beta \nonumber \\
&=& \frac{1}{D(D-1)} \left| \int e^{-iE_\alpha t}  R_1(E_\alpha) dE_\alpha \right|^2 .
\label{Eq:R1quad}
\ea
In accordance with Fig.~\ref{fig:C}~(b), we use that the width of the Gaussian density of states is approximately the same as the width of the LDOS, $\Gamma_\text{DOS} \sim \Gamma$, and write
\be
R_1(E)= \frac{D}{ \sqrt{2\pi}\Gamma {\cal N}}\exp\left(-\frac{E^2}{2\Gamma^2}\right).
\label{Eq:Gauss_SM}
\ee
In addition, the spectrum is bounded~\cite{Tavora2016,Tavora2017}  between energies $E_\textrm{min}$ and $E_\textrm{max}$, which explains the normalization factor,
\be
{\cal N} = \frac{1}{2}  \left[ \textrm{erf}\left(\frac{E_\textrm{max}}{\sqrt{2}\Gamma}\right)-\textrm{erf}\left(\frac{E_\textrm{min}}{\sqrt{2}\Gamma}\right) \right].
\ee
Plugging Eq.~(\ref{Eq:Gauss_SM}) into Eq.~(\ref{Eq:R1quad}) gives
\be
\frac{1}{D(D-1)}\left|\int_{E_\textrm{min}}^{E_\textrm{max}} dE e^{-iEt} R_1(E)\right|^2=\frac{D}{D-1}\frac{e^{-\Gamma^2t^2}}{4 {\cal N}^2} {\cal F}(t),
\label{Eq:R1quad2}
\ee
where
\be
{\cal F}(t) = \left|\textrm{erf}\left(\frac{E_\textrm{max}+it\Gamma^2}{\sqrt{2}\Gamma}\right)-\textrm{erf}\left(\frac{E_\textrm{min}+it\Gamma^2}{\sqrt{2}\Gamma}\right)\right|^2 .
\ee
In the above, $\textrm{erf}$ is the error function. 

For very short times, $t\ll 1/\Gamma$, Eq.~(\ref{Eq:R1quad2}) leads to the universal quadratic decay of the survival probability $1 - \Gamma^2 t^2$. This is followed by a true  Gaussian behavior,  $\exp (-\Gamma^2t^2)$, as expected from the Fourier transform of a Gaussian energy distribution~\cite{Santos2012PRL,*Santos2012PRE,Torres2014PRA,Torres2014NJP,Torres2014PRE,Torres2014PRAb}.

For long times,  Eq.~(\ref{Eq:R1quad2})  can be written as~\cite{Tavora2016,Tavora2017},
\ba
&& \frac{D}{D-1}\frac{1}{2 \pi{\cal N}^2\Gamma^2t^2}\left[\exp \left( -\frac{E_\textrm{max}^2}{\Gamma^2} \right) +\exp \left( -\frac{E_\textrm{min}^2}{\Gamma^2} \right) \right. \nonumber \\
&&\left. \!-2 \exp \left( -\frac{E_\textrm{max}^2+E_\textrm{min}^2}{2\Gamma^2} \right)
\cos[(E_\textrm{max} \!-\! E_\textrm{min})t]\right].
\label{Eq:resultR1_SM}
\ea
Since the cosine term averages to zero at large times, we are left with
\be
\frac{D}{D-1}\frac{1}{2 \pi {\cal N}^2 \Gamma^2t^2}\left[ 
\exp \left( -\frac{E_\textrm{max}^2}{\Gamma^2} \right) +\exp \left( -\frac{E_\textrm{min}^2}{\Gamma^2} \right) 
\right],
\label{Eq:long_SM}
\ee
which shows that, later in time, a power-law decay $\propto t^{-2}$ develops.

\subsection{Correlation hole}

Let us now go back to Eq.~(\ref{Eq:R2_SM}) and compute the Fourier transform of the second term,
\ba
&&-\frac{(D-2)!}{D!}  \times \nonumber \\
&&  \int e^{-iEt} \delta(E-E_\alpha+E_\beta)T_2(E_\alpha,E_\beta) dE dE_\alpha dE_\beta.\nonumber
\ea
For full random matrices, following Ref.~\cite{MehtaBook}, one writes the energies in terms of the mean level spacing, $\mu=1/R_1(0)$, introducing the variables $\epsilon_{\alpha,\beta}\equiv E_{\alpha,\beta}/\mu$. In the limit $D\rightarrow\infty$, one has
\ba
&&- \frac{(D-2)!}{D!} \int e^{-i(E_\alpha-E_\beta)t} T_2(E_\alpha,E_\beta) \, dE_\alpha \, dE_\beta=\nonumber\\
&&- \frac{(D-2)!}{D!} \int e^{-i \mu   (\epsilon_\alpha-\epsilon_\beta)t} Y_2(\epsilon_\alpha,\epsilon_\beta)d\epsilon_\alpha d\epsilon_\beta,
\label{Eq:T2}
\ea
where $Y_2(\epsilon_\alpha,\epsilon_\beta) = \mu^2 T_2(E_\alpha,E_\beta)$. In the bulk of the spectrum, the cluster function is translation-invariant, {\em i.e.} $Y_2(\epsilon_\alpha,\epsilon_\beta)=Y_2(r)$, with $r=\left|\epsilon_\alpha-\epsilon_\beta\right|$. This is not true if $E_\alpha$ or $E_\beta$ are close to the boundaries of the spectrum, but such anomalous contributions are negligible for large $D$. Taking into account the change in variables, the Fourier transform of $Y_2(r)$ gives~\cite{MehtaBook}
\be
- \frac{(D-2)!}{D!} \int D e^{-i r \mu t} Y_2(r)dr=\frac{1}{D-1} b_2\left(\frac{\mu t}{2\pi}\right) ,
\label{Eq:resultR2_SM}
\ee
where
\ba
b_2(t) &=& [1-2t + t \ln(1+2 t)] \Theta (1- t) \\
&+& \{t \ln [ (2 t+1)/(2 t -1) ] -1 \} \Theta(t-1),\nonumber
\ea
is the two-level form factor presented in Eq.~(\ref{eq:b2}).

For chaotic noninteracting disordered quantum systems in more than two dimensions, spectral correlations are analogous to those found in random matrices for energy separations $|E_\alpha-E_\beta|\ll E_\textrm{Th}$, with $E_\textrm{Th} \gg s$ being the Thouless energy~\cite{Altshuler1986,Guhr1998}. 
The same is also true for chaotic interacting systems~\cite{Bertrand2016}. Furthermore, it is known that $T_2(E_\alpha,E_\beta)$ is exponentially small for $|E_\alpha-E_\beta| \gg s$, so the same procedure to get the $b_2$ function described above holds for realistic chaotic systems as well, the only difference in this case is that the mean level spacing comes from the Gaussian distribution, $\mu=\sqrt{2 \pi} \Gamma/D$,

Plugging Eqs.~(\ref{Eq:R1quad2}) and~(\ref{Eq:resultR2_SM}) into Eq.~(\ref{eq:SM_delta}), and this one back into  Eq.~(\ref{Eq:FourierApp}), one obtains the final expression of Eq.~(\ref{eq:PSrealistic}), that is,
\be
\left<P_S(t)\right> =\frac{1-\overline{P_S}}{ (D-1)}\left[\frac{ D e^{-\Gamma^2t^2}}{4 {\cal N}^2} {\cal F}(t) -b_2\left(\frac{\Gamma t}{\sqrt{2\pi}D}\right)\right] + \overline{P_S} .
\ee

\bibliography{biblioThouless}

\begin{thebibliography}{74}%
\makeatletter
\providecommand \@ifxundefined [1]{%
 \@ifx{#1\undefined}
}%
\providecommand \@ifnum [1]{%
 \ifnum #1\expandafter \@firstoftwo
 \else \expandafter \@secondoftwo
 \fi
}%
\providecommand \@ifx [1]{%
 \ifx #1\expandafter \@firstoftwo
 \else \expandafter \@secondoftwo
 \fi
}%
\providecommand \natexlab [1]{#1}%
\providecommand \enquote  [1]{``#1''}%
\providecommand \bibnamefont  [1]{#1}%
\providecommand \bibfnamefont [1]{#1}%
\providecommand \citenamefont [1]{#1}%
\providecommand \href@noop [0]{\@secondoftwo}%
\providecommand \href [0]{\begingroup \@sanitize@url \@href}%
\providecommand \@href[1]{\@@startlink{#1}\@@href}%
\providecommand \@@href[1]{\endgroup#1\@@endlink}%
\providecommand \@sanitize@url [0]{\catcode `\\12\catcode `\$12\catcode
  `\&12\catcode `\#12\catcode `\^12\catcode `\_12\catcode `\%12\relax}%
\providecommand \@@startlink[1]{}%
\providecommand \@@endlink[0]{}%
\providecommand \url  [0]{\begingroup\@sanitize@url \@url }%
\providecommand \@url [1]{\endgroup\@href {#1}{\urlprefix }}%
\providecommand \urlprefix  [0]{URL }%
\providecommand \Eprint [0]{\href }%
\providecommand \doibase [0]{http://dx.doi.org/}%
\providecommand \selectlanguage [0]{\@gobble}%
\providecommand \bibinfo  [0]{\@secondoftwo}%
\providecommand \bibfield  [0]{\@secondoftwo}%
\providecommand \translation [1]{[#1]}%
\providecommand \BibitemOpen [0]{}%
\providecommand \bibitemStop [0]{}%
\providecommand \bibitemNoStop [0]{.\EOS\space}%
\providecommand \EOS [0]{\spacefactor3000\relax}%
\providecommand \BibitemShut  [1]{\csname bibitem#1\endcsname}%
\let\auto@bib@innerbib\@empty
\bibitem [{\citenamefont {Gavish}\ and\ \citenamefont
  {Castin}(2005)}]{Gavish2005}%
  \BibitemOpen
  \bibfield  {author} {\bibinfo {author} {\bibfnamefont {U.}~\bibnamefont
  {Gavish}}\ and\ \bibinfo {author} {\bibfnamefont {Y.}~\bibnamefont
  {Castin}},\ }\bibfield  {title} {\enquote {\bibinfo {title} {Matter-wave
  localization in disordered cold atom lattices},}\ }\href {\doibase
  10.1103/PhysRevLett.95.020401} {\bibfield  {journal} {\bibinfo  {journal}
  {Phys. Rev. Lett.}\ }\textbf {\bibinfo {volume} {95}},\ \bibinfo {pages}
  {020401} (\bibinfo {year} {2005})}\BibitemShut {NoStop}%
\bibitem [{\citenamefont {Bloch}\ \emph {et~al.}(2008)\citenamefont {Bloch},
  \citenamefont {Dalibard},\ and\ \citenamefont {Zwerger}}]{Bloch2008}%
  \BibitemOpen
  \bibfield  {author} {\bibinfo {author} {\bibfnamefont {I.}~\bibnamefont
  {Bloch}}, \bibinfo {author} {\bibfnamefont {J.}~\bibnamefont {Dalibard}}, \
  and\ \bibinfo {author} {\bibfnamefont {W.}~\bibnamefont {Zwerger}},\
  }\bibfield  {title} {\enquote {\bibinfo {title} {Many-body physics with
  ultracold gases},}\ }\href {\doibase 10.1103/RevModPhys.80.885} {\bibfield
  {journal} {\bibinfo  {journal} {Rev. Mod. Phys.}\ }\textbf {\bibinfo {volume}
  {80}},\ \bibinfo {pages} {885--964} (\bibinfo {year} {2008})}\BibitemShut
  {NoStop}%
\bibitem [{\citenamefont {Gadway}\ \emph {et~al.}(2010)\citenamefont {Gadway},
  \citenamefont {Pertot}, \citenamefont {Reimann},\ and\ \citenamefont
  {Schneble}}]{Gadway2010}%
  \BibitemOpen
  \bibfield  {author} {\bibinfo {author} {\bibfnamefont {B.}~\bibnamefont
  {Gadway}}, \bibinfo {author} {\bibfnamefont {D.}~\bibnamefont {Pertot}},
  \bibinfo {author} {\bibfnamefont {R.}~\bibnamefont {Reimann}}, \ and\
  \bibinfo {author} {\bibfnamefont {D.}~\bibnamefont {Schneble}},\ }\bibfield
  {title} {\enquote {\bibinfo {title} {Superfluidity of interacting bosonic
  mixtures in optical lattices},}\ }\href {\doibase
  10.1103/PhysRevLett.105.045303} {\bibfield  {journal} {\bibinfo  {journal}
  {Phys. Rev. Lett.}\ }\textbf {\bibinfo {volume} {105}},\ \bibinfo {pages}
  {045303} (\bibinfo {year} {2010})}\BibitemShut {NoStop}%
\bibitem [{\citenamefont {Bloch}\ \emph {et~al.}(2012)\citenamefont {Bloch},
  \citenamefont {Dalibard},\ and\ \citenamefont {Nascimb\`ene}}]{Bloch2012}%
  \BibitemOpen
  \bibfield  {author} {\bibinfo {author} {\bibfnamefont {I.}~\bibnamefont
  {Bloch}}, \bibinfo {author} {\bibfnamefont {J.}~\bibnamefont {Dalibard}}, \
  and\ \bibinfo {author} {\bibfnamefont {S.}~\bibnamefont {Nascimb\`ene}},\
  }\bibfield  {title} {\enquote {\bibinfo {title} {Quantum simulations with
  ultracold quantum gases},}\ }\href@noop {} {\bibfield  {journal} {\bibinfo
  {journal} {Nat. Phys.}\ }\textbf {\bibinfo {volume} {8}},\ \bibinfo {pages}
  {267--276} (\bibinfo {year} {2012})}\BibitemShut {NoStop}%
\bibitem [{\citenamefont {Jurcevic}\ \emph {et~al.}(2014)\citenamefont
  {Jurcevic}, \citenamefont {Lanyon}, \citenamefont {Hauke}, \citenamefont
  {Hempel}, \citenamefont {Zoller}, \citenamefont {Blatt},\ and\ \citenamefont
  {Roos}}]{Jurcevic2014}%
  \BibitemOpen
  \bibfield  {author} {\bibinfo {author} {\bibfnamefont {P.}~\bibnamefont
  {Jurcevic}}, \bibinfo {author} {\bibfnamefont {B.~P.}\ \bibnamefont
  {Lanyon}}, \bibinfo {author} {\bibfnamefont {P.}~\bibnamefont {Hauke}},
  \bibinfo {author} {\bibfnamefont {C.}~\bibnamefont {Hempel}}, \bibinfo
  {author} {\bibfnamefont {P.}~\bibnamefont {Zoller}}, \bibinfo {author}
  {\bibfnamefont {R.}~\bibnamefont {Blatt}}, \ and\ \bibinfo {author}
  {\bibfnamefont {C.~F.}\ \bibnamefont {Roos}},\ }\bibfield  {title} {\enquote
  {\bibinfo {title} {Quasiparticle engineering and entanglement propagation in
  a quantum many-body system},}\ }\href@noop {} {\bibfield  {journal} {\bibinfo
   {journal} {Nature}\ }\textbf {\bibinfo {volume} {511}},\ \bibinfo {pages}
  {202--205} (\bibinfo {year} {2014})}\BibitemShut {NoStop}%
\bibitem [{\citenamefont {Richerme}\ \emph {et~al.}(2014)\citenamefont
  {Richerme}, \citenamefont {Gong}, \citenamefont {Lee}, \citenamefont {Senko},
  \citenamefont {Smith}, \citenamefont {Foss-Feig}, \citenamefont {Michalakis},
  \citenamefont {Gorshkov},\ and\ \citenamefont {Monroe}}]{Richerme2014}%
  \BibitemOpen
  \bibfield  {author} {\bibinfo {author} {\bibfnamefont {P.}~\bibnamefont
  {Richerme}}, \bibinfo {author} {\bibfnamefont {Z.-X.}\ \bibnamefont {Gong}},
  \bibinfo {author} {\bibfnamefont {A.}~\bibnamefont {Lee}}, \bibinfo {author}
  {\bibfnamefont {Cr.}\ \bibnamefont {Senko}}, \bibinfo {author} {\bibfnamefont
  {J.}~\bibnamefont {Smith}}, \bibinfo {author} {\bibfnamefont
  {M.}~\bibnamefont {Foss-Feig}}, \bibinfo {author} {\bibfnamefont
  {S.}~\bibnamefont {Michalakis}}, \bibinfo {author} {\bibfnamefont {A.~V.}\
  \bibnamefont {Gorshkov}}, \ and\ \bibinfo {author} {\bibfnamefont
  {C.}~\bibnamefont {Monroe}},\ }\bibfield  {title} {\enquote {\bibinfo {title}
  {Non-local propagation of correlations in quantum systems with long-range
  interactions},}\ }\href@noop {} {\bibfield  {journal} {\bibinfo  {journal}
  {Nature}\ }\textbf {\bibinfo {volume} {511}},\ \bibinfo {pages} {198--201}
  (\bibinfo {year} {2014})}\BibitemShut {NoStop}%
\bibitem [{\citenamefont {Schreiber}\ \emph {et~al.}(2015)\citenamefont
  {Schreiber}, \citenamefont {Hodgman}, \citenamefont {Bordia}, \citenamefont
  {L{\"u}schen}, \citenamefont {Fischer}, \citenamefont {Vosk}, \citenamefont
  {Altman}, \citenamefont {Schneider},\ and\ \citenamefont
  {Bloch}}]{Schreiber2015}%
  \BibitemOpen
  \bibfield  {author} {\bibinfo {author} {\bibfnamefont {M.}~\bibnamefont
  {Schreiber}}, \bibinfo {author} {\bibfnamefont {S.~S.}\ \bibnamefont
  {Hodgman}}, \bibinfo {author} {\bibfnamefont {Pr.}\ \bibnamefont {Bordia}},
  \bibinfo {author} {\bibfnamefont {H.~P.}\ \bibnamefont {L{\"u}schen}},
  \bibinfo {author} {\bibfnamefont {M.~H.}\ \bibnamefont {Fischer}}, \bibinfo
  {author} {\bibfnamefont {R.}~\bibnamefont {Vosk}}, \bibinfo {author}
  {\bibfnamefont {E.}~\bibnamefont {Altman}}, \bibinfo {author} {\bibfnamefont
  {U.}~\bibnamefont {Schneider}}, \ and\ \bibinfo {author} {\bibfnamefont
  {I.}~\bibnamefont {Bloch}},\ }\bibfield  {title} {\enquote {\bibinfo {title}
  {Observation of many-body localization of interacting fermions in a
  quasirandom optical lattice},}\ }\href {\doibase 10.1126/science.aaa7432}
  {\bibfield  {journal} {\bibinfo  {journal} {Science}\ }\textbf {\bibinfo
  {volume} {349}},\ \bibinfo {pages} {842--845} (\bibinfo {year}
  {2015})}\BibitemShut {NoStop}%
\bibitem [{\citenamefont {G\"arttner}\ \emph {et~al.}(2017)\citenamefont
  {G\"arttner}, \citenamefont {Bohnet}, \citenamefont {Safavi-Naini},
  \citenamefont {Wall}, \citenamefont {Bollinger},\ and\ \citenamefont
  {Rey}}]{Garttner2017}%
  \BibitemOpen
  \bibfield  {author} {\bibinfo {author} {\bibfnamefont {M.}~\bibnamefont
  {G\"arttner}}, \bibinfo {author} {\bibfnamefont {J.~G.}\ \bibnamefont
  {Bohnet}}, \bibinfo {author} {\bibfnamefont {A.}~\bibnamefont
  {Safavi-Naini}}, \bibinfo {author} {\bibfnamefont {M.~L.}\ \bibnamefont
  {Wall}}, \bibinfo {author} {\bibfnamefont {J.~J.}\ \bibnamefont {Bollinger}},
  \ and\ \bibinfo {author} {\bibfnamefont {A.~M.}\ \bibnamefont {Rey}},\
  }\bibfield  {title} {\enquote {\bibinfo {title} {Measuring out-of-time-order
  correlations and multiple quantum spectra in a trapped-ion quantum magnet},}\
  }\href {http://dx.doi.org/10.1038/nphys4119} {\bibfield  {journal} {\bibinfo
  {journal} {Nat. Phys.}\ }\textbf {\bibinfo {volume} {13}},\ \bibinfo {pages}
  {781 -- 786} (\bibinfo {year} {2017})}\BibitemShut {NoStop}%
\bibitem [{\citenamefont {Wei}\ \emph {et~al.}(2018)\citenamefont {Wei},
  \citenamefont {Ramanathan},\ and\ \citenamefont {Cappellaro}}]{Wei2018}%
  \BibitemOpen
  \bibfield  {author} {\bibinfo {author} {\bibfnamefont {K.~X.}\ \bibnamefont
  {Wei}}, \bibinfo {author} {\bibfnamefont {C.}~\bibnamefont {Ramanathan}}, \
  and\ \bibinfo {author} {\bibfnamefont {P.}~\bibnamefont {Cappellaro}},\
  }\bibfield  {title} {\enquote {\bibinfo {title} {Exploring localization in
  nuclear spin chains},}\ }\href {\doibase 10.1103/PhysRevLett.120.070501}
  {\bibfield  {journal} {\bibinfo  {journal} {Phys. Rev. Lett.}\ }\textbf
  {\bibinfo {volume} {120}},\ \bibinfo {pages} {070501} (\bibinfo {year}
  {2018})}\BibitemShut {NoStop}%
\bibitem [{\citenamefont {Gogolin}\ and\ \citenamefont
  {Eisert}(2016)}]{Gogolin2016}%
  \BibitemOpen
  \bibfield  {author} {\bibinfo {author} {\bibfnamefont {C.}~\bibnamefont
  {Gogolin}}\ and\ \bibinfo {author} {\bibfnamefont {J.}~\bibnamefont
  {Eisert}},\ }\bibfield  {title} {\enquote {\bibinfo {title} {Equilibration,
  thermalisation, and the emergence of statistical mechanics in closed quantum
  systems},}\ }\href {http://stacks.iop.org/0034-4885/79/i=5/a=056001}
  {\bibfield  {journal} {\bibinfo  {journal} {Rep. Prog. Phys.}\ }\textbf
  {\bibinfo {volume} {79}},\ \bibinfo {pages} {056001} (\bibinfo {year}
  {2016})}\BibitemShut {NoStop}%
\bibitem [{\citenamefont {Borgonovi}\ \emph {et~al.}(2016)\citenamefont
  {Borgonovi}, \citenamefont {Izrailev}, \citenamefont {Santos},\ and\
  \citenamefont {Zelevinsky}}]{Borgonovi2016}%
  \BibitemOpen
  \bibfield  {author} {\bibinfo {author} {\bibfnamefont {F.}~\bibnamefont
  {Borgonovi}}, \bibinfo {author} {\bibfnamefont {F.~M.}\ \bibnamefont
  {Izrailev}}, \bibinfo {author} {\bibfnamefont {L.~F.}\ \bibnamefont
  {Santos}}, \ and\ \bibinfo {author} {\bibfnamefont {V.~G.}\ \bibnamefont
  {Zelevinsky}},\ }\bibfield  {title} {\enquote {\bibinfo {title} {Quantum
  chaos and thermalization in isolated systems of interacting particles},}\
  }\href {\doibase 10.1016/j.physrep.2016.02.005} {\bibfield  {journal}
  {\bibinfo  {journal} {Phys. Rep.}\ }\textbf {\bibinfo {volume} {626}},\
  \bibinfo {pages} {1} (\bibinfo {year} {2016})}\BibitemShut {NoStop}%
\bibitem [{\citenamefont {Alessio}\ \emph {et~al.}(2016)\citenamefont
  {Alessio}, \citenamefont {Kafri}, \citenamefont {Polkovnikov},\ and\
  \citenamefont {Rigol}}]{Dalessio2016}%
  \BibitemOpen
  \bibfield  {author} {\bibinfo {author} {\bibfnamefont {L.~D'}\ \bibnamefont
  {Alessio}}, \bibinfo {author} {\bibfnamefont {Y.}~\bibnamefont {Kafri}},
  \bibinfo {author} {\bibfnamefont {A.}~\bibnamefont {Polkovnikov}}, \ and\
  \bibinfo {author} {\bibfnamefont {M.}~\bibnamefont {Rigol}},\ }\bibfield
  {title} {\enquote {\bibinfo {title} {From quantum chaos and eigenstate
  thermalization to statistical mechanics and thermodynamics},}\ }\href
  {\doibase 10.1080/00018732.2016.1198134} {\bibfield  {journal} {\bibinfo
  {journal} {Adv. Phys.}\ }\textbf {\bibinfo {volume} {65}},\ \bibinfo {pages}
  {239--362} (\bibinfo {year} {2016})}\BibitemShut {NoStop}%
\bibitem [{\citenamefont {Dymarsky}({\natexlab{a}})}]{DymarskyARXIV}%
  \BibitemOpen
  \bibfield  {author} {\bibinfo {author} {\bibfnamefont {A.}~\bibnamefont
  {Dymarsky}},\ }\href@noop {} {\enquote {\bibinfo {title} {Bound on eigenstate
  thermalization from transport},}\ } ({\natexlab{a}}),\ \bibinfo {note}
  {arXiv:1804.08626}\BibitemShut {NoStop}%
\bibitem [{\citenamefont {Reimann}(2018{\natexlab{a}})}]{Reimann2018a}%
  \BibitemOpen
  \bibfield  {author} {\bibinfo {author} {\bibfnamefont {P.}~\bibnamefont
  {Reimann}},\ }\bibfield  {title} {\enquote {\bibinfo {title} {Dynamical
  typicality approach to eigenstate thermalization},}\ }\href {\doibase
  10.1103/PhysRevLett.120.230601} {\bibfield  {journal} {\bibinfo  {journal}
  {Phys. Rev. Lett.}\ }\textbf {\bibinfo {volume} {120}},\ \bibinfo {pages}
  {230601} (\bibinfo {year} {2018}{\natexlab{a}})}\BibitemShut {NoStop}%
\bibitem [{\citenamefont {Reimann}(2018{\natexlab{b}})}]{Reimann2018b}%
  \BibitemOpen
  \bibfield  {author} {\bibinfo {author} {\bibfnamefont {P.}~\bibnamefont
  {Reimann}},\ }\bibfield  {title} {\enquote {\bibinfo {title} {Dynamical
  typicality of isolated many-body quantum systems},}\ }\href {\doibase
  10.1103/PhysRevE.97.062129} {\bibfield  {journal} {\bibinfo  {journal} {Phys.
  Rev. E}\ }\textbf {\bibinfo {volume} {97}},\ \bibinfo {pages} {062129}
  (\bibinfo {year} {2018}{\natexlab{b}})}\BibitemShut {NoStop}%
\bibitem [{\citenamefont {Santos}\ \emph {et~al.}(2005)\citenamefont {Santos},
  \citenamefont {Dykman}, \citenamefont {Shapiro},\ and\ \citenamefont
  {Izrailev}}]{Santos2005}%
  \BibitemOpen
  \bibfield  {author} {\bibinfo {author} {\bibfnamefont {L.~F.}\ \bibnamefont
  {Santos}}, \bibinfo {author} {\bibfnamefont {M.~I.}\ \bibnamefont {Dykman}},
  \bibinfo {author} {\bibfnamefont {M.}~\bibnamefont {Shapiro}}, \ and\
  \bibinfo {author} {\bibfnamefont {F.~M.}\ \bibnamefont {Izrailev}},\
  }\bibfield  {title} {\enquote {\bibinfo {title} {Strong many-particle
  localization and quantum computing with perpetually coupled qubits},}\ }\href
  {\doibase 10.1103/PhysRevA.71.012317} {\bibfield  {journal} {\bibinfo
  {journal} {Phys. Rev. A}\ }\textbf {\bibinfo {volume} {71}},\ \bibinfo
  {pages} {012317} (\bibinfo {year} {2005})}\BibitemShut {NoStop}%
\bibitem [{\citenamefont {Nandkishore}\ and\ \citenamefont
  {Huse}(2015)}]{Nandkishore2015}%
  \BibitemOpen
  \bibfield  {author} {\bibinfo {author} {\bibfnamefont {R.}~\bibnamefont
  {Nandkishore}}\ and\ \bibinfo {author} {\bibfnamefont {D.A.}\ \bibnamefont
  {Huse}},\ }\bibfield  {title} {\enquote {\bibinfo {title} {Many-body
  localization and thermalization in quantum statistical mechanics},}\
  }\href@noop {} {\bibfield  {journal} {\bibinfo  {journal} {Annu. Rev.
  Condens. Matter Phys.}\ }\textbf {\bibinfo {volume} {6}},\ \bibinfo {pages}
  {15} (\bibinfo {year} {2015})}\BibitemShut {NoStop}%
\bibitem [{\citenamefont {Luitz}\ and\ \citenamefont {Lev}(2017)}]{Luitz2017}%
  \BibitemOpen
  \bibfield  {author} {\bibinfo {author} {\bibfnamefont {D.}~\bibnamefont
  {Luitz}}\ and\ \bibinfo {author} {\bibfnamefont {Y.~Bar}\ \bibnamefont
  {Lev}},\ }\bibfield  {title} {\enquote {\bibinfo {title} {The ergodic side of
  the many-body localization transition},}\ }\href {\doibase
  10.1002/andp.201600350} {\bibfield  {journal} {\bibinfo  {journal} {Ann.
  Phys.(Berlin)}\ }\textbf {\bibinfo {volume} {529}},\ \bibinfo {pages}
  {1600350} (\bibinfo {year} {2017})}\BibitemShut {NoStop}%
\bibitem [{\citenamefont {Serbyn}\ \emph {et~al.}(2015)\citenamefont {Serbyn},
  \citenamefont {Papi\ifmmode~\acute{c}\else \'{c}\fi{}},\ and\ \citenamefont
  {Abanin}}]{Serbyn2015}%
  \BibitemOpen
  \bibfield  {author} {\bibinfo {author} {\bibfnamefont {M.}~\bibnamefont
  {Serbyn}}, \bibinfo {author} {\bibfnamefont {Z.}~\bibnamefont
  {Papi\ifmmode~\acute{c}\else \'{c}\fi{}}}, \ and\ \bibinfo {author}
  {\bibfnamefont {D.~A.}\ \bibnamefont {Abanin}},\ }\bibfield  {title}
  {\enquote {\bibinfo {title} {Criterion for many-body
  localization-delocalization phase transition},}\ }\href {\doibase
  10.1103/PhysRevX.5.041047} {\bibfield  {journal} {\bibinfo  {journal} {Phys.
  Rev. X}\ }\textbf {\bibinfo {volume} {5}},\ \bibinfo {pages} {041047}
  (\bibinfo {year} {2015})}\BibitemShut {NoStop}%
\bibitem [{\citenamefont {Serbyn}\ \emph {et~al.}(2017)\citenamefont {Serbyn},
  \citenamefont {Papi\ifmmode~\acute{c}\else \'{c}\fi{}},\ and\ \citenamefont
  {Abanin}}]{Serbyn2017}%
  \BibitemOpen
  \bibfield  {author} {\bibinfo {author} {\bibfnamefont {M.}~\bibnamefont
  {Serbyn}}, \bibinfo {author} {\bibfnamefont {Z.}~\bibnamefont
  {Papi\ifmmode~\acute{c}\else \'{c}\fi{}}}, \ and\ \bibinfo {author}
  {\bibfnamefont {D.~A.}\ \bibnamefont {Abanin}},\ }\bibfield  {title}
  {\enquote {\bibinfo {title} {Thouless energy and multifractality across the
  many-body localization transition},}\ }\href {\doibase
  10.1103/PhysRevB.96.104201} {\bibfield  {journal} {\bibinfo  {journal} {Phys.
  Rev. B}\ }\textbf {\bibinfo {volume} {96}},\ \bibinfo {pages} {104201}
  (\bibinfo {year} {2017})}\BibitemShut {NoStop}%
\bibitem [{\citenamefont {Varma}\ \emph {et~al.}(2017)\citenamefont {Varma},
  \citenamefont {Lerose}, \citenamefont {Pietracaprina}, \citenamefont
  {Goold},\ and\ \citenamefont {Scardicchio}}]{Varma2017}%
  \BibitemOpen
  \bibfield  {author} {\bibinfo {author} {\bibfnamefont {V.~K.}\ \bibnamefont
  {Varma}}, \bibinfo {author} {\bibfnamefont {A.}~\bibnamefont {Lerose}},
  \bibinfo {author} {\bibfnamefont {F.}~\bibnamefont {Pietracaprina}}, \bibinfo
  {author} {\bibfnamefont {J.}~\bibnamefont {Goold}}, \ and\ \bibinfo {author}
  {\bibfnamefont {A.}~\bibnamefont {Scardicchio}},\ }\bibfield  {title}
  {\enquote {\bibinfo {title} {Energy diffusion in the ergodic phase of a many
  body localizable spin chain},}\ }\href {\doibase 10.1088/1742-5468/aa668b}
  {\bibfield  {journal} {\bibinfo  {journal} {J. Stat. Mech.: Th. Exp.}\
  }\textbf {\bibinfo {volume} {2017}},\ \bibinfo {pages} {053101} (\bibinfo
  {year} {2017})}\BibitemShut {NoStop}%
\bibitem [{\citenamefont {Scaffidi}\ and\ \citenamefont
  {Altman}()}]{ScaffidiARXIV}%
  \BibitemOpen
  \bibfield  {author} {\bibinfo {author} {\bibfnamefont {T.}~\bibnamefont
  {Scaffidi}}\ and\ \bibinfo {author} {\bibfnamefont {E.}~\bibnamefont
  {Altman}},\ }\href@noop {} {\enquote {\bibinfo {title} {Semiclassical theory
  of many-body quantum chaos and its bound},}\ }\bibinfo {note}
  {ArXiv:1711.04768}\BibitemShut {NoStop}%
\bibitem [{\citenamefont {Rammensee}\ \emph {et~al.}(2018)\citenamefont
  {Rammensee}, \citenamefont {Urbina},\ and\ \citenamefont
  {Richter}}]{Rammensee2018}%
  \BibitemOpen
  \bibfield  {author} {\bibinfo {author} {\bibfnamefont {J.}~\bibnamefont
  {Rammensee}}, \bibinfo {author} {\bibfnamefont {J.~D.}\ \bibnamefont
  {Urbina}}, \ and\ \bibinfo {author} {\bibfnamefont {K.}~\bibnamefont
  {Richter}},\ }\bibfield  {title} {\enquote {\bibinfo {title} {Many-body
  quantum interference and the saturation of out-of-time-order correlators},}\
  }\href {\doibase 10.1103/PhysRevLett.121.124101} {\bibfield  {journal}
  {\bibinfo  {journal} {Phys. Rev. Lett.}\ }\textbf {\bibinfo {volume} {121}},\
  \bibinfo {pages} {124101} (\bibinfo {year} {2018})}\BibitemShut {NoStop}%
\bibitem [{\citenamefont {Cotler}\ \emph {et~al.}(2017)\citenamefont {Cotler},
  \citenamefont {Hunter-Jones}, \citenamefont {Liu},\ and\ \citenamefont
  {Yoshida}}]{Cotler2017GUE}%
  \BibitemOpen
  \bibfield  {author} {\bibinfo {author} {\bibfnamefont {J.}~\bibnamefont
  {Cotler}}, \bibinfo {author} {\bibfnamefont {N.}~\bibnamefont
  {Hunter-Jones}}, \bibinfo {author} {\bibfnamefont {J.}~\bibnamefont {Liu}}, \
  and\ \bibinfo {author} {\bibfnamefont {B.}~\bibnamefont {Yoshida}},\
  }\bibfield  {title} {\enquote {\bibinfo {title} {Chaos, complexity, and
  random matrices},}\ }\href {\doibase 10.1007/JHEP11(2017)048} {\bibfield
  {journal} {\bibinfo  {journal} {J. High Energy Phys.}\ }\textbf {\bibinfo
  {volume} {2017}},\ \bibinfo {pages} {48} (\bibinfo {year}
  {2017})}\BibitemShut {NoStop}%
\bibitem [{\citenamefont {Gharibyan}\ \emph {et~al.}(2018)\citenamefont
  {Gharibyan}, \citenamefont {Hanada}, \citenamefont {Shenker},\ and\
  \citenamefont {Tezuka}}]{Gharibyan2018}%
  \BibitemOpen
  \bibfield  {author} {\bibinfo {author} {\bibfnamefont {H.}~\bibnamefont
  {Gharibyan}}, \bibinfo {author} {\bibfnamefont {M.}~\bibnamefont {Hanada}},
  \bibinfo {author} {\bibfnamefont {S.~H.}\ \bibnamefont {Shenker}}, \ and\
  \bibinfo {author} {\bibfnamefont {M.}~\bibnamefont {Tezuka}},\ }\bibfield
  {title} {\enquote {\bibinfo {title} {Onset of random matrix behavior in
  scrambling systems},}\ }\href {\doibase 10.1007/JHEP07(2018)124} {\bibfield
  {journal} {\bibinfo  {journal} {Journal of High Energy Physics}\ }\textbf
  {\bibinfo {volume} {2018}},\ \bibinfo {pages} {124} (\bibinfo {year}
  {2018})}\BibitemShut {NoStop}%
\bibitem [{\citenamefont {Nosaka}\ \emph {et~al.}(2018)\citenamefont {Nosaka},
  \citenamefont {Rosa},\ and\ \citenamefont {Yoon}}]{Nosaka2018}%
  \BibitemOpen
  \bibfield  {author} {\bibinfo {author} {\bibfnamefont {T.}~\bibnamefont
  {Nosaka}}, \bibinfo {author} {\bibfnamefont {D.}~\bibnamefont {Rosa}}, \ and\
  \bibinfo {author} {\bibfnamefont {J.}~\bibnamefont {Yoon}},\ }\bibfield
  {title} {\enquote {\bibinfo {title} {The {T}houless time for mass-deformed
  {SYK}},}\ }\href {\doibase 10.1007/JHEP09(2018)041} {\bibfield  {journal}
  {\bibinfo  {journal} {J. High Energy Phys.}\ }\textbf {\bibinfo {volume}
  {2018}},\ \bibinfo {pages} {41} (\bibinfo {year} {2018})}\BibitemShut
  {NoStop}%
\bibitem [{\citenamefont {Chan}\ \emph {et~al.}(2018)\citenamefont {Chan},
  \citenamefont {De~Luca},\ and\ \citenamefont {Chalker}}]{Chan2018}%
  \BibitemOpen
  \bibfield  {author} {\bibinfo {author} {\bibfnamefont {A.}~\bibnamefont
  {Chan}}, \bibinfo {author} {\bibfnamefont {A.}~\bibnamefont {De~Luca}}, \
  and\ \bibinfo {author} {\bibfnamefont {J.~T.}\ \bibnamefont {Chalker}},\
  }\bibfield  {title} {\enquote {\bibinfo {title} {Spectral statistics in
  spatially extended chaotic quantum many-body systems},}\ }\href {\doibase
  10.1103/PhysRevLett.121.060601} {\bibfield  {journal} {\bibinfo  {journal}
  {Phys. Rev. Lett.}\ }\textbf {\bibinfo {volume} {121}},\ \bibinfo {pages}
  {060601} (\bibinfo {year} {2018})}\BibitemShut {NoStop}%
\bibitem [{\citenamefont {F.~Borgonovi}()}]{Borgonovi2019}%
  \BibitemOpen
  \bibfield  {author} {\bibinfo {author} {\bibfnamefont {L.~F.~Santos}\
  \bibnamefont {F.~Borgonovi}, \bibfnamefont {F.~M.~Izrailev}},\ }\href@noop {}
  {\enquote {\bibinfo {title} {Timescales in the quench dynamics of many-body
  quantum systems: Participation ratio vs out-of-time ordered correlator},}\
  }\bibinfo {note} {ArXiv:1903.09175}\BibitemShut {NoStop}%
\bibitem [{\citenamefont {Borgonovi}\ \emph {et~al.}()\citenamefont
  {Borgonovi}, \citenamefont {Izrailev},\ and\ \citenamefont
  {Santos}}]{BorgonoviARXIV}%
  \BibitemOpen
  \bibfield  {author} {\bibinfo {author} {\bibfnamefont {F.}~\bibnamefont
  {Borgonovi}}, \bibinfo {author} {\bibfnamefont {F.~M.}\ \bibnamefont
  {Izrailev}}, \ and\ \bibinfo {author} {\bibfnamefont {L.~F.}\ \bibnamefont
  {Santos}},\ }\href@noop {} {\enquote {\bibinfo {title} {Exponentially fast
  dynamics in the {F}ock space of chaotic many-body systems},}\ }\bibinfo
  {note} {ArXiv:1802.08265}\BibitemShut {NoStop}%
\bibitem [{\citenamefont {Peres}(1984)}]{Peres1984}%
  \BibitemOpen
  \bibfield  {author} {\bibinfo {author} {\bibfnamefont {A.}~\bibnamefont
  {Peres}},\ }\bibfield  {title} {\enquote {\bibinfo {title} {Stability of
  quantum motion in chaotic and regular systems},}\ }\href@noop {} {\bibfield
  {journal} {\bibinfo  {journal} {Phys. Rev. A}\ }\textbf {\bibinfo {volume}
  {30}},\ \bibinfo {pages} {1610--1615} (\bibinfo {year} {1984})}\BibitemShut
  {NoStop}%
\bibitem [{\citenamefont {Deutsch}(1991)}]{Deutsch1991}%
  \BibitemOpen
  \bibfield  {author} {\bibinfo {author} {\bibfnamefont {J.~M.}\ \bibnamefont
  {Deutsch}},\ }\bibfield  {title} {\enquote {\bibinfo {title} {Quantum
  statistical mechanics in a closed system},}\ }\href@noop {} {\bibfield
  {journal} {\bibinfo  {journal} {Phys. Rev. A}\ }\textbf {\bibinfo {volume}
  {43}},\ \bibinfo {pages} {2046} (\bibinfo {year} {1991})}\BibitemShut
  {NoStop}%
\bibitem [{\citenamefont {Srednicki}(1996)}]{Srednicki1996}%
  \BibitemOpen
  \bibfield  {author} {\bibinfo {author} {\bibfnamefont {M.}~\bibnamefont
  {Srednicki}},\ }\bibfield  {title} {\enquote {\bibinfo {title} {Thermal
  fluctuations in quantized chaotic systems},}\ }\href@noop {} {\bibfield
  {journal} {\bibinfo  {journal} {J. Phys. A}\ }\textbf {\bibinfo {volume}
  {29}},\ \bibinfo {pages} {L75--L79} (\bibinfo {year} {1996})}\BibitemShut
  {NoStop}%
\bibitem [{\citenamefont {Reimann}(2008)}]{Reimann2008}%
  \BibitemOpen
  \bibfield  {author} {\bibinfo {author} {\bibfnamefont {P.}~\bibnamefont
  {Reimann}},\ }\bibfield  {title} {\enquote {\bibinfo {title} {Foundation of
  statistical mechanics under experimentally realistic conditions},}\
  }\href@noop {} {\bibfield  {journal} {\bibinfo  {journal} {Phys. Rev. Lett.}\
  }\textbf {\bibinfo {volume} {101}},\ \bibinfo {pages} {190403} (\bibinfo
  {year} {2008})}\BibitemShut {NoStop}%
\bibitem [{\citenamefont {Short}(2011)}]{Short2011}%
  \BibitemOpen
  \bibfield  {author} {\bibinfo {author} {\bibfnamefont {A.~J.}\ \bibnamefont
  {Short}},\ }\bibfield  {title} {\enquote {\bibinfo {title} {Equilibration of
  quantum systems and subsystems},}\ }\href@noop {} {\bibfield  {journal}
  {\bibinfo  {journal} {New J. Phys.}\ }\textbf {\bibinfo {volume} {13}},\
  \bibinfo {pages} {053009} (\bibinfo {year} {2011})}\BibitemShut {NoStop}%
\bibitem [{\citenamefont {Short}\ and\ \citenamefont
  {Farrelly}(2012)}]{Short2012}%
  \BibitemOpen
  \bibfield  {author} {\bibinfo {author} {\bibfnamefont {A.~J.}\ \bibnamefont
  {Short}}\ and\ \bibinfo {author} {\bibfnamefont {T.~C.}\ \bibnamefont
  {Farrelly}},\ }\bibfield  {title} {\enquote {\bibinfo {title} {Quantum
  equilibration in finite time},}\ }\href@noop {} {\bibfield  {journal}
  {\bibinfo  {journal} {New J. Phys.}\ }\textbf {\bibinfo {volume} {14}},\
  \bibinfo {pages} {013063} (\bibinfo {year} {2012})}\BibitemShut {NoStop}%
\bibitem [{\citenamefont {He}\ \emph {et~al.}(2013)\citenamefont {He},
  \citenamefont {Santos}, \citenamefont {Wright},\ and\ \citenamefont
  {Rigol}}]{HeSantos2013}%
  \BibitemOpen
  \bibfield  {author} {\bibinfo {author} {\bibfnamefont {K.}~\bibnamefont
  {He}}, \bibinfo {author} {\bibfnamefont {L.~F.}\ \bibnamefont {Santos}},
  \bibinfo {author} {\bibfnamefont {T.~M.}\ \bibnamefont {Wright}}, \ and\
  \bibinfo {author} {\bibfnamefont {M.}~\bibnamefont {Rigol}},\ }\bibfield
  {title} {\enquote {\bibinfo {title} {Single-particle and many-body analyses
  of a quasiperiodic integrable system after a quench},}\ }\href {\doibase
  10.1103/PhysRevA.87.063637} {\bibfield  {journal} {\bibinfo  {journal} {Phys.
  Rev. A}\ }\textbf {\bibinfo {volume} {87}},\ \bibinfo {pages} {063637}
  (\bibinfo {year} {2013})}\BibitemShut {NoStop}%
\bibitem [{\citenamefont {Zangara}\ \emph {et~al.}(2013)\citenamefont
  {Zangara}, \citenamefont {Dente}, \citenamefont {Torres-Herrera},
  \citenamefont {Pastawski}, \citenamefont {Iucci},\ and\ \citenamefont
  {Santos}}]{Zangara2013}%
  \BibitemOpen
  \bibfield  {author} {\bibinfo {author} {\bibfnamefont {P.~R.}\ \bibnamefont
  {Zangara}}, \bibinfo {author} {\bibfnamefont {A.~D.}\ \bibnamefont {Dente}},
  \bibinfo {author} {\bibfnamefont {E.~J.}\ \bibnamefont {Torres-Herrera}},
  \bibinfo {author} {\bibfnamefont {H.~M.}\ \bibnamefont {Pastawski}}, \bibinfo
  {author} {\bibfnamefont {A.}~\bibnamefont {Iucci}}, \ and\ \bibinfo {author}
  {\bibfnamefont {L.~F.}\ \bibnamefont {Santos}},\ }\bibfield  {title}
  {\enquote {\bibinfo {title} {Time fluctuations in isolated quantum systems of
  interacting particles},}\ }\href@noop {} {\bibfield  {journal} {\bibinfo
  {journal} {Phys. Rev. E}\ }\textbf {\bibinfo {volume} {88}},\ \bibinfo
  {pages} {032913} (\bibinfo {year} {2013})}\BibitemShut {NoStop}%
\bibitem [{\citenamefont {Monnai}(2013)}]{Monnai2013}%
  \BibitemOpen
  \bibfield  {author} {\bibinfo {author} {\bibfnamefont {T.}~\bibnamefont
  {Monnai}},\ }\bibfield  {title} {\enquote {\bibinfo {title} {Generic
  evaluation of relaxation time for quantum many-body systems: Analysis of the
  system size dependence},}\ }\href {\doibase 10.7566/JPSJ.82.044006}
  {\bibfield  {journal} {\bibinfo  {journal} {J. Phys. Soc. Jpn.}\ }\textbf
  {\bibinfo {volume} {82}},\ \bibinfo {pages} {044006} (\bibinfo {year}
  {2013})}\BibitemShut {NoStop}%
\bibitem [{\citenamefont {Goldstein}\ \emph {et~al.}(2013)\citenamefont
  {Goldstein}, \citenamefont {Hara},\ and\ \citenamefont
  {Tasaki}}]{Goldstein2013}%
  \BibitemOpen
  \bibfield  {author} {\bibinfo {author} {\bibfnamefont {S.}~\bibnamefont
  {Goldstein}}, \bibinfo {author} {\bibfnamefont {T.}~\bibnamefont {Hara}}, \
  and\ \bibinfo {author} {\bibfnamefont {H.}~\bibnamefont {Tasaki}},\
  }\bibfield  {title} {\enquote {\bibinfo {title} {Time scales in the approach
  to equilibrium of macroscopic quantum systems},}\ }\href {\doibase
  10.1103/PhysRevLett.111.140401} {\bibfield  {journal} {\bibinfo  {journal}
  {Phys. Rev. Lett.}\ }\textbf {\bibinfo {volume} {111}},\ \bibinfo {pages}
  {140401} (\bibinfo {year} {2013})}\BibitemShut {NoStop}%
\bibitem [{\citenamefont {Malabarba}\ \emph {et~al.}(2014)\citenamefont
  {Malabarba}, \citenamefont {Garc\'{\i}a-Pintos}, \citenamefont {Linden},
  \citenamefont {Farrelly},\ and\ \citenamefont {Short}}]{Malabarba2014}%
  \BibitemOpen
  \bibfield  {author} {\bibinfo {author} {\bibfnamefont {A.~S.~L.}\
  \bibnamefont {Malabarba}}, \bibinfo {author} {\bibfnamefont {L.~P.}\
  \bibnamefont {Garc\'{\i}a-Pintos}}, \bibinfo {author} {\bibfnamefont
  {N.}~\bibnamefont {Linden}}, \bibinfo {author} {\bibfnamefont {T.~C.}\
  \bibnamefont {Farrelly}}, \ and\ \bibinfo {author} {\bibfnamefont {A.~J.}\
  \bibnamefont {Short}},\ }\bibfield  {title} {\enquote {\bibinfo {title}
  {Quantum systems equilibrate rapidly for most observables},}\ }\href
  {\doibase 10.1103/PhysRevE.90.012121} {\bibfield  {journal} {\bibinfo
  {journal} {Phys. Rev. E}\ }\textbf {\bibinfo {volume} {90}},\ \bibinfo
  {pages} {012121} (\bibinfo {year} {2014})}\BibitemShut {NoStop}%
\bibitem [{\citenamefont {Goldstein}\ \emph {et~al.}(2015)\citenamefont
  {Goldstein}, \citenamefont {Hara},\ and\ \citenamefont
  {Tasaki}}]{Goldstein2015}%
  \BibitemOpen
  \bibfield  {author} {\bibinfo {author} {\bibfnamefont {S.}~\bibnamefont
  {Goldstein}}, \bibinfo {author} {\bibfnamefont {T.}~\bibnamefont {Hara}}, \
  and\ \bibinfo {author} {\bibfnamefont {H.}~\bibnamefont {Tasaki}},\
  }\bibfield  {title} {\enquote {\bibinfo {title} {Extremely quick
  thermalization in a macroscopic quantum system for a typical nonequilibrium
  subspace},}\ }\href {http://stacks.iop.org/1367-2630/17/i=4/a=045002}
  {\bibfield  {journal} {\bibinfo  {journal} {New J. Phys.}\ }\textbf {\bibinfo
  {volume} {17}},\ \bibinfo {pages} {045002} (\bibinfo {year}
  {2015})}\BibitemShut {NoStop}%
\bibitem [{\citenamefont {Reimann}(2016)}]{Reimann2016}%
  \BibitemOpen
  \bibfield  {author} {\bibinfo {author} {\bibfnamefont {P.}~\bibnamefont
  {Reimann}},\ }\bibfield  {title} {\enquote {\bibinfo {title} {Typical fast
  thermalization processes in closed many-body systems},}\ }\href {\doibase
  http://dx.doi.org/10.1038/ncomms10821} {\bibfield  {journal} {\bibinfo
  {journal} {Nat. Comm.}\ }\textbf {\bibinfo {volume} {7}},\ \bibinfo {pages}
  {10821} (\bibinfo {year} {2016})}\BibitemShut {NoStop}%
\bibitem [{\citenamefont {Garc\'{\i}a-Pintos}\ \emph
  {et~al.}(2017)\citenamefont {Garc\'{\i}a-Pintos}, \citenamefont {Linden},
  \citenamefont {Malabarba}, \citenamefont {Short},\ and\ \citenamefont
  {Winter}}]{Pintos2017}%
  \BibitemOpen
  \bibfield  {author} {\bibinfo {author} {\bibfnamefont {L.~P.}\ \bibnamefont
  {Garc\'{\i}a-Pintos}}, \bibinfo {author} {\bibfnamefont {N.}~\bibnamefont
  {Linden}}, \bibinfo {author} {\bibfnamefont {A.~S.~L.}\ \bibnamefont
  {Malabarba}}, \bibinfo {author} {\bibfnamefont {A.~J.}\ \bibnamefont
  {Short}}, \ and\ \bibinfo {author} {\bibfnamefont {A.}~\bibnamefont
  {Winter}},\ }\bibfield  {title} {\enquote {\bibinfo {title} {Equilibration
  time scales of physically relevant observables},}\ }\href {\doibase
  10.1103/PhysRevX.7.031027} {\bibfield  {journal} {\bibinfo  {journal} {Phys.
  Rev. X}\ }\textbf {\bibinfo {volume} {7}},\ \bibinfo {pages} {031027}
  (\bibinfo {year} {2017})}\BibitemShut {NoStop}%
\bibitem [{\citenamefont {de~Oliveira}\ \emph {et~al.}(2018)\citenamefont
  {de~Oliveira}, \citenamefont {Charalambous}, \citenamefont {Jonathan},
  \citenamefont {Lewenstein},\ and\ \citenamefont {Riera}}]{Oliveira2018}%
  \BibitemOpen
  \bibfield  {author} {\bibinfo {author} {\bibfnamefont {T.~R.}\ \bibnamefont
  {de~Oliveira}}, \bibinfo {author} {\bibfnamefont {C.}~\bibnamefont
  {Charalambous}}, \bibinfo {author} {\bibfnamefont {D.}~\bibnamefont
  {Jonathan}}, \bibinfo {author} {\bibfnamefont {M.}~\bibnamefont
  {Lewenstein}}, \ and\ \bibinfo {author} {\bibfnamefont {A.}~\bibnamefont
  {Riera}},\ }\bibfield  {title} {\enquote {\bibinfo {title} {Equilibration
  time scales in closed many-body quantum systems},}\ }\href
  {http://stacks.iop.org/1367-2630/20/i=3/a=033032} {\bibfield  {journal}
  {\bibinfo  {journal} {New J. Phys.}\ }\textbf {\bibinfo {volume} {20}},\
  \bibinfo {pages} {033032} (\bibinfo {year} {2018})}\BibitemShut {NoStop}%
\bibitem [{\citenamefont {Dymarsky}({\natexlab{b}})}]{DymarskyARXIVThouless}%
  \BibitemOpen
  \bibfield  {author} {\bibinfo {author} {\bibfnamefont {A.}~\bibnamefont
  {Dymarsky}},\ }\href@noop {} {\enquote {\bibinfo {title} {Mechanism of slow
  equilibration of isolated quantum systems},}\ } ({\natexlab{b}}),\ \bibinfo
  {note} {arXiv:1806.04187}\BibitemShut {NoStop}%
\bibitem [{\citenamefont {Guhr}\ \emph {et~al.}(1998)\citenamefont {Guhr},
  \citenamefont {Mueller-Gr\"oeling},\ and\ \citenamefont
  {Weidenm\"uller}}]{Guhr1998}%
  \BibitemOpen
  \bibfield  {author} {\bibinfo {author} {\bibfnamefont {T.}~\bibnamefont
  {Guhr}}, \bibinfo {author} {\bibfnamefont {A.}~\bibnamefont
  {Mueller-Gr\"oeling}}, \ and\ \bibinfo {author} {\bibfnamefont {H.~A.}\
  \bibnamefont {Weidenm\"uller}},\ }\bibfield  {title} {\enquote {\bibinfo
  {title} {Random matrix theories in quantum physics: Common concepts},}\
  }\href@noop {} {\bibfield  {journal} {\bibinfo  {journal} {Phys. Rep.}\
  }\textbf {\bibinfo {volume} {299}},\ \bibinfo {pages} {189} (\bibinfo {year}
  {1998})}\BibitemShut {NoStop}%
\bibitem [{\citenamefont {Mehta}(1991)}]{MehtaBook}%
  \BibitemOpen
  \bibfield  {author} {\bibinfo {author} {\bibfnamefont {M.~L.}\ \bibnamefont
  {Mehta}},\ }\href@noop {} {\emph {\bibinfo {title} {Random Matrices}}}\
  (\bibinfo  {publisher} {Academic Press},\ \bibinfo {address} {Boston, USA},\
  \bibinfo {year} {1991})\BibitemShut {NoStop}%
\bibitem [{\citenamefont {Torres-Herrera}\ \emph {et~al.}(2018)\citenamefont
  {Torres-Herrera}, \citenamefont {Garc\'{\i}a-Garc\'{\i}a},\ and\
  \citenamefont {Santos}}]{Torres2018}%
  \BibitemOpen
  \bibfield  {author} {\bibinfo {author} {\bibfnamefont {E.~J.}\ \bibnamefont
  {Torres-Herrera}}, \bibinfo {author} {\bibfnamefont {Antonio~M.}\
  \bibnamefont {Garc\'{\i}a-Garc\'{\i}a}}, \ and\ \bibinfo {author}
  {\bibfnamefont {Lea~F.}\ \bibnamefont {Santos}},\ }\bibfield  {title}
  {\enquote {\bibinfo {title} {Generic dynamical features of quenched
  interacting quantum systems: Survival probability, density imbalance, and
  out-of-time-ordered correlator},}\ }\href {\doibase
  10.1103/PhysRevB.97.060303} {\bibfield  {journal} {\bibinfo  {journal} {Phys.
  Rev. B}\ }\textbf {\bibinfo {volume} {97}},\ \bibinfo {pages} {060303}
  (\bibinfo {year} {2018})}\BibitemShut {NoStop}%
\bibitem [{\citenamefont {Leviandier}\ \emph {et~al.}(1986)\citenamefont
  {Leviandier}, \citenamefont {Lombardi}, \citenamefont {Jost},\ and\
  \citenamefont {Pique}}]{Leviandier1986}%
  \BibitemOpen
  \bibfield  {author} {\bibinfo {author} {\bibfnamefont {L.}~\bibnamefont
  {Leviandier}}, \bibinfo {author} {\bibfnamefont {M.}~\bibnamefont
  {Lombardi}}, \bibinfo {author} {\bibfnamefont {R.}~\bibnamefont {Jost}}, \
  and\ \bibinfo {author} {\bibfnamefont {J.~P.}\ \bibnamefont {Pique}},\
  }\bibfield  {title} {\enquote {\bibinfo {title} {Fourier transform: A tool to
  measure statistical level properties in very complex spectra},}\ }\href
  {\doibase 10.1103/PhysRevLett.56.2449} {\bibfield  {journal} {\bibinfo
  {journal} {Phys. Rev. Lett.}\ }\textbf {\bibinfo {volume} {56}},\ \bibinfo
  {pages} {2449--2452} (\bibinfo {year} {1986})}\BibitemShut {NoStop}%
\bibitem [{\citenamefont {Guhr}\ and\ \citenamefont
  {Weidenm\"uller}(1990)}]{Guhr1990}%
  \BibitemOpen
  \bibfield  {author} {\bibinfo {author} {\bibfnamefont {T.}~\bibnamefont
  {Guhr}}\ and\ \bibinfo {author} {\bibfnamefont {H.A.}\ \bibnamefont
  {Weidenm\"uller}},\ }\bibfield  {title} {\enquote {\bibinfo {title}
  {Correlations in anticrossing spectra and scattering theory. analytical
  aspects},}\ }\href {\doibase http://dx.doi.org/10.1016/0301-0104(90)90003-R}
  {\bibfield  {journal} {\bibinfo  {journal} {Chem. Phys.}\ }\textbf {\bibinfo
  {volume} {146}},\ \bibinfo {pages} {21 -- 38} (\bibinfo {year}
  {1990})}\BibitemShut {NoStop}%
\bibitem [{\citenamefont {Wilkie}\ and\ \citenamefont
  {Brumer}(1991)}]{Wilkie1991}%
  \BibitemOpen
  \bibfield  {author} {\bibinfo {author} {\bibfnamefont {J.}~\bibnamefont
  {Wilkie}}\ and\ \bibinfo {author} {\bibfnamefont {P.}~\bibnamefont
  {Brumer}},\ }\bibfield  {title} {\enquote {\bibinfo {title} {Time-dependent
  manifestations of quantum chaos},}\ }\href {\doibase
  10.1103/PhysRevLett.67.1185} {\bibfield  {journal} {\bibinfo  {journal}
  {Phys. Rev. Lett.}\ }\textbf {\bibinfo {volume} {67}},\ \bibinfo {pages}
  {1185--1188} (\bibinfo {year} {1991})}\BibitemShut {NoStop}%
\bibitem [{\citenamefont {Alhassid}\ and\ \citenamefont
  {Levine}(1992)}]{Alhassid1992}%
  \BibitemOpen
  \bibfield  {author} {\bibinfo {author} {\bibfnamefont {Y.}~\bibnamefont
  {Alhassid}}\ and\ \bibinfo {author} {\bibfnamefont {R.~D.}\ \bibnamefont
  {Levine}},\ }\bibfield  {title} {\enquote {\bibinfo {title} {Spectral
  autocorrelation function in the statistical theory of energy levels},}\
  }\href {\doibase 10.1103/PhysRevA.46.4650} {\bibfield  {journal} {\bibinfo
  {journal} {Phys. Rev. A}\ }\textbf {\bibinfo {volume} {46}},\ \bibinfo
  {pages} {4650--4653} (\bibinfo {year} {1992})}\BibitemShut {NoStop}%
\bibitem [{\citenamefont {Gorin}\ and\ \citenamefont
  {Seligman}(2002)}]{Gorin2002}%
  \BibitemOpen
  \bibfield  {author} {\bibinfo {author} {\bibfnamefont {T.}~\bibnamefont
  {Gorin}}\ and\ \bibinfo {author} {\bibfnamefont {T.~H.}\ \bibnamefont
  {Seligman}},\ }\bibfield  {title} {\enquote {\bibinfo {title} {Signatures of
  the correlation hole in total and partial cross sections},}\ }\href {\doibase
  10.1103/PhysRevE.65.026214} {\bibfield  {journal} {\bibinfo  {journal} {Phys.
  Rev. E}\ }\textbf {\bibinfo {volume} {65}},\ \bibinfo {pages} {026214}
  (\bibinfo {year} {2002})}\BibitemShut {NoStop}%
\bibitem [{\citenamefont {Torres-Herrera}\ and\ \citenamefont
  {Santos}(2017{\natexlab{a}})}]{Torres2017}%
  \BibitemOpen
  \bibfield  {author} {\bibinfo {author} {\bibfnamefont {E.~J.}\ \bibnamefont
  {Torres-Herrera}}\ and\ \bibinfo {author} {\bibfnamefont {Lea~F.}\
  \bibnamefont {Santos}},\ }\bibfield  {title} {\enquote {\bibinfo {title}
  {Extended nonergodic states in disordered many-body quantum systems},}\
  }\href {\doibase 10.1002/andp.201600284} {\bibfield  {journal} {\bibinfo
  {journal} {Ann. Phys. (Berlin)}\ }\textbf {\bibinfo {volume} {529}},\
  \bibinfo {pages} {1600284} (\bibinfo {year}
  {2017}{\natexlab{a}})}\BibitemShut {NoStop}%
\bibitem [{\citenamefont {Torres-Herrera}\ and\ \citenamefont
  {Santos}(2017{\natexlab{b}})}]{Torres2017Philo}%
  \BibitemOpen
  \bibfield  {author} {\bibinfo {author} {\bibfnamefont {E.~J.}\ \bibnamefont
  {Torres-Herrera}}\ and\ \bibinfo {author} {\bibfnamefont {Lea~F.}\
  \bibnamefont {Santos}},\ }\bibfield  {title} {\enquote {\bibinfo {title}
  {Dynamical manifestations of quantum chaos: correlation hole and bulge},}\
  }\href {\doibase 10.1098/rsta.2016.0434} {\bibfield  {journal} {\bibinfo
  {journal} {Philos. Trans. R. Soc. London A}\ }\textbf {\bibinfo {volume}
  {375}},\ \bibinfo {pages} {20160434} (\bibinfo {year}
  {2017}{\natexlab{b}})}\BibitemShut {NoStop}%
\bibitem [{\citenamefont {Thouless}(1974)}]{Thouless1974}%
  \BibitemOpen
  \bibfield  {author} {\bibinfo {author} {\bibfnamefont {D.J.}\ \bibnamefont
  {Thouless}},\ }\bibfield  {title} {\enquote {\bibinfo {title} {Electrons in
  disordered systems and the theory of localization},}\ }\href {\doibase
  https://doi.org/10.1016/0370-1573(74)90029-5} {\bibfield  {journal} {\bibinfo
   {journal} {Phys. Rep.}\ }\textbf {\bibinfo {volume} {13}},\ \bibinfo {pages}
  {93 -- 142} (\bibinfo {year} {1974})}\BibitemShut {NoStop}%
\bibitem [{\citenamefont {Al'tshuler}\ and\ \citenamefont
  {Shklovskii}(1986)}]{Altshuler1986}%
  \BibitemOpen
  \bibfield  {author} {\bibinfo {author} {\bibfnamefont {B.~L.}\ \bibnamefont
  {Al'tshuler}}\ and\ \bibinfo {author} {\bibfnamefont {B.~I.}\ \bibnamefont
  {Shklovskii}},\ }\bibfield  {title} {\enquote {\bibinfo {title} {Repulsion of
  energy levels and conductivity of small metal samples},}\ }\href@noop {}
  {\bibfield  {journal} {\bibinfo  {journal} {Zh. Eksp. Teor. Fiz.}\ }\textbf
  {\bibinfo {volume} {91}},\ \bibinfo {pages} {220} (\bibinfo {year} {1986})},\
  \bibinfo {note} {[Sov. Phys. JETP {\bf 64}, 127 (1986)]}\BibitemShut
  {NoStop}%
\bibitem [{\citenamefont {Al'tshuler}\ \emph {et~al.}(1988)\citenamefont
  {Al'tshuler}, \citenamefont {Zharekeshev}, \citenamefont {Kotochigova},\ and\
  \citenamefont {Shklovskii}}]{Altshuler1988}%
  \BibitemOpen
  \bibfield  {author} {\bibinfo {author} {\bibfnamefont {B.~L.}\ \bibnamefont
  {Al'tshuler}}, \bibinfo {author} {\bibfnamefont {I.~Kh.}\ \bibnamefont
  {Zharekeshev}}, \bibinfo {author} {\bibfnamefont {S.~A.}\ \bibnamefont
  {Kotochigova}}, \ and\ \bibinfo {author} {\bibfnamefont {B.~I.}\ \bibnamefont
  {Shklovskii}},\ }\bibfield  {title} {\enquote {\bibinfo {title} {Repulsion
  between energy levels and the metal-insulator transition},}\ }\href@noop {}
  {\bibfield  {journal} {\bibinfo  {journal} {Zh. Eksp. Teor. Fiz.}\ }\textbf
  {\bibinfo {volume} {94}},\ \bibinfo {pages} {343} (\bibinfo {year} {1988})},\
  \bibinfo {note} {[Sov. Phys. JETP {\bf 67}, 625 (1988)]}\BibitemShut
  {NoStop}%
\bibitem [{\citenamefont {Bertrand}\ and\ \citenamefont
  {Garc\'{\i}a-Garc\'{\i}a}(2016)}]{Bertrand2016}%
  \BibitemOpen
  \bibfield  {author} {\bibinfo {author} {\bibfnamefont {C.~L.}\ \bibnamefont
  {Bertrand}}\ and\ \bibinfo {author} {\bibfnamefont {A.~M.}\ \bibnamefont
  {Garc\'{\i}a-Garc\'{\i}a}},\ }\bibfield  {title} {\enquote {\bibinfo {title}
  {Anomalous thouless energy and critical statistics on the metallic side of
  the many-body localization transition},}\ }\href {\doibase
  10.1103/PhysRevB.94.144201} {\bibfield  {journal} {\bibinfo  {journal} {Phys.
  Rev. B}\ }\textbf {\bibinfo {volume} {94}},\ \bibinfo {pages} {144201}
  (\bibinfo {year} {2016})}\BibitemShut {NoStop}%
\bibitem [{\citenamefont {Torres-Herrera}\ and\ \citenamefont
  {Santos}(2015)}]{Torres2015}%
  \BibitemOpen
  \bibfield  {author} {\bibinfo {author} {\bibfnamefont {E.~J.}\ \bibnamefont
  {Torres-Herrera}}\ and\ \bibinfo {author} {\bibfnamefont {Lea~F.}\
  \bibnamefont {Santos}},\ }\bibfield  {title} {\enquote {\bibinfo {title}
  {Dynamics at the many-body localization transition},}\ }\href {\doibase
  10.1103/PhysRevB.92.014208} {\bibfield  {journal} {\bibinfo  {journal} {Phys.
  Rev. B}\ }\textbf {\bibinfo {volume} {92}},\ \bibinfo {pages} {014208}
  (\bibinfo {year} {2015})}\BibitemShut {NoStop}%
\bibitem [{\citenamefont {Santos}\ and\ \citenamefont
  {Torres-Herrera}(2017)}]{SantosTorres2017AIP}%
  \BibitemOpen
  \bibfield  {author} {\bibinfo {author} {\bibfnamefont {L.~F.}\ \bibnamefont
  {Santos}}\ and\ \bibinfo {author} {\bibfnamefont {E.~J.}\ \bibnamefont
  {Torres-Herrera}},\ }\bibfield  {title} {\enquote {\bibinfo {title}
  {Analytical expressions for the evolution of many-body quantum systems
  quenched far from equilibrium},}\ }\href {\doibase 10.1063/1.5016140}
  {\bibfield  {journal} {\bibinfo  {journal} {AIP Conference Proceedings}\
  }\textbf {\bibinfo {volume} {1912}},\ \bibinfo {pages} {020015} (\bibinfo
  {year} {2017})}\BibitemShut {NoStop}%
\bibitem [{\citenamefont {Torres-Herrera}\ \emph {et~al.}(2015)\citenamefont
  {Torres-Herrera}, \citenamefont {Kollmar},\ and\ \citenamefont
  {Santos}}]{TorresKollmar2015}%
  \BibitemOpen
  \bibfield  {author} {\bibinfo {author} {\bibfnamefont {E.~J.}\ \bibnamefont
  {Torres-Herrera}}, \bibinfo {author} {\bibfnamefont {D.}~\bibnamefont
  {Kollmar}}, \ and\ \bibinfo {author} {\bibfnamefont {L.~F.}\ \bibnamefont
  {Santos}},\ }\bibfield  {title} {\enquote {\bibinfo {title} {Relaxation and
  thermalization of isolated many-body quantum systems},}\ }\href@noop {}
  {\bibfield  {journal} {\bibinfo  {journal} {Phys. Scr. T}\ }\textbf {\bibinfo
  {volume} {165}},\ \bibinfo {pages} {014018} (\bibinfo {year}
  {2015})}\BibitemShut {NoStop}%
\bibitem [{\citenamefont {Brody}\ \emph {et~al.}(1981)\citenamefont {Brody},
  \citenamefont {Flores}, \citenamefont {French}, \citenamefont {Mello},
  \citenamefont {Pandey},\ and\ \citenamefont {Wong}}]{Brody1981}%
  \BibitemOpen
  \bibfield  {author} {\bibinfo {author} {\bibfnamefont {T.~A.}\ \bibnamefont
  {Brody}}, \bibinfo {author} {\bibfnamefont {J.}~\bibnamefont {Flores}},
  \bibinfo {author} {\bibfnamefont {J.~B.}\ \bibnamefont {French}}, \bibinfo
  {author} {\bibfnamefont {P.~A.}\ \bibnamefont {Mello}}, \bibinfo {author}
  {\bibfnamefont {A.}~\bibnamefont {Pandey}}, \ and\ \bibinfo {author}
  {\bibfnamefont {S.~S.~M.}\ \bibnamefont {Wong}},\ }\bibfield  {title}
  {\enquote {\bibinfo {title} {Random-matrix physics: spectrum and strength
  fluctuations},}\ }\href {\doibase 10.1103/RevModPhys.53.385} {\bibfield
  {journal} {\bibinfo  {journal} {Rev. Mod. Phys.}\ }\textbf {\bibinfo {volume}
  {53}},\ \bibinfo {pages} {385} (\bibinfo {year} {1981})}\BibitemShut
  {NoStop}%
\bibitem [{\citenamefont {Torres-Herrera}\ and\ \citenamefont
  {Santos}(2014{\natexlab{a}})}]{Torres2014PRA}%
  \BibitemOpen
  \bibfield  {author} {\bibinfo {author} {\bibfnamefont {E.~J.}\ \bibnamefont
  {Torres-Herrera}}\ and\ \bibinfo {author} {\bibfnamefont {Lea~F.}\
  \bibnamefont {Santos}},\ }\bibfield  {title} {\enquote {\bibinfo {title}
  {Quench dynamics of isolated many-body quantum systems},}\ }\href {\doibase
  10.1103/PhysRevA.89.043620} {\bibfield  {journal} {\bibinfo  {journal} {Phys.
  Rev. A}\ }\textbf {\bibinfo {volume} {89}},\ \bibinfo {pages} {043620}
  (\bibinfo {year} {2014}{\natexlab{a}})}\BibitemShut {NoStop}%
\bibitem [{\citenamefont {Torres-Herrera}\ \emph {et~al.}(2014)\citenamefont
  {Torres-Herrera}, \citenamefont {Vyas},\ and\ \citenamefont
  {Santos}}]{Torres2014NJP}%
  \BibitemOpen
  \bibfield  {author} {\bibinfo {author} {\bibfnamefont {E.~J.}\ \bibnamefont
  {Torres-Herrera}}, \bibinfo {author} {\bibfnamefont {M.}~\bibnamefont
  {Vyas}}, \ and\ \bibinfo {author} {\bibfnamefont {Lea~F.}\ \bibnamefont
  {Santos}},\ }\bibfield  {title} {\enquote {\bibinfo {title} {General features
  of the relaxation dynamics of interacting quantum systems},}\ }\href@noop {}
  {\bibfield  {journal} {\bibinfo  {journal} {New J. Phys.}\ }\textbf {\bibinfo
  {volume} {16}},\ \bibinfo {pages} {063010} (\bibinfo {year}
  {2014})}\BibitemShut {NoStop}%
\bibitem [{\citenamefont {T\'avora}\ \emph {et~al.}(2016)\citenamefont
  {T\'avora}, \citenamefont {Torres-Herrera},\ and\ \citenamefont
  {Santos}}]{Tavora2016}%
  \BibitemOpen
  \bibfield  {author} {\bibinfo {author} {\bibfnamefont {M.}~\bibnamefont
  {T\'avora}}, \bibinfo {author} {\bibfnamefont {E.~J.}\ \bibnamefont
  {Torres-Herrera}}, \ and\ \bibinfo {author} {\bibfnamefont {L.~F.}\
  \bibnamefont {Santos}},\ }\bibfield  {title} {\enquote {\bibinfo {title}
  {Inevitable power-law behavior of isolated many-body quantum systems and how
  it anticipates thermalization},}\ }\href {\doibase
  10.1103/PhysRevA.94.041603} {\bibfield  {journal} {\bibinfo  {journal} {Phys.
  Rev. A}\ }\textbf {\bibinfo {volume} {94}},\ \bibinfo {pages} {041603}
  (\bibinfo {year} {2016})}\BibitemShut {NoStop}%
\bibitem [{\citenamefont {T\'avora}\ \emph {et~al.}(2017)\citenamefont
  {T\'avora}, \citenamefont {Torres-Herrera},\ and\ \citenamefont
  {Santos}}]{Tavora2017}%
  \BibitemOpen
  \bibfield  {author} {\bibinfo {author} {\bibfnamefont {M.}~\bibnamefont
  {T\'avora}}, \bibinfo {author} {\bibfnamefont {E.~J.}\ \bibnamefont
  {Torres-Herrera}}, \ and\ \bibinfo {author} {\bibfnamefont {L.~F.}\
  \bibnamefont {Santos}},\ }\bibfield  {title} {\enquote {\bibinfo {title}
  {Power-law decay exponents: A dynamical criterion for predicting
  thermalization},}\ }\href {\doibase 10.1103/PhysRevA.95.013604} {\bibfield
  {journal} {\bibinfo  {journal} {Phys. Rev. A}\ }\textbf {\bibinfo {volume}
  {95}},\ \bibinfo {pages} {013604} (\bibinfo {year} {2017})}\BibitemShut
  {NoStop}%
\bibitem [{\citenamefont {Torres-Herrera}\ and\ \citenamefont
  {Santos}()}]{TorresARXIV}%
  \BibitemOpen
  \bibfield  {author} {\bibinfo {author} {\bibfnamefont {E.~J.}\ \bibnamefont
  {Torres-Herrera}}\ and\ \bibinfo {author} {\bibfnamefont {L.~F.}\
  \bibnamefont {Santos}},\ }\href@noop {} {\enquote {\bibinfo {title}
  {Signatures of chaos and thermalization in the dynamics of many-body quantum
  systems},}\ }\bibinfo {note} {ArXiv:1804.06401}\BibitemShut {NoStop}%
\bibitem [{\citenamefont {Schiulaz}\ \emph {et~al.}(2018)\citenamefont
  {Schiulaz}, \citenamefont {T\'avora},\ and\ \citenamefont
  {Santos}}]{Schiulaz2018}%
  \BibitemOpen
  \bibfield  {author} {\bibinfo {author} {\bibfnamefont {M.}~\bibnamefont
  {Schiulaz}}, \bibinfo {author} {\bibfnamefont {M.}~\bibnamefont {T\'avora}},
  \ and\ \bibinfo {author} {\bibfnamefont {L.~F.}\ \bibnamefont {Santos}},\
  }\bibfield  {title} {\enquote {\bibinfo {title} {From few- to many-body
  quantum systems},}\ }\href@noop {} {\bibfield  {journal} {\bibinfo  {journal}
  {Quantum Sci. Technol.}\ }\textbf {\bibinfo {volume} {3}},\ \bibinfo {pages}
  {044006} (\bibinfo {year} {2018})}\BibitemShut {NoStop}%
\bibitem [{\citenamefont {Torres-Herrera}\ and\ \citenamefont
  {Santos}(2014{\natexlab{b}})}]{Torres2014PRE}%
  \BibitemOpen
  \bibfield  {author} {\bibinfo {author} {\bibfnamefont {E.~J.}\ \bibnamefont
  {Torres-Herrera}}\ and\ \bibinfo {author} {\bibfnamefont {Lea~F.}\
  \bibnamefont {Santos}},\ }\bibfield  {title} {\enquote {\bibinfo {title}
  {Local quenches with global effects in interacting quantum systems},}\ }\href
  {\doibase 10.1103/PhysRevE.89.062110} {\bibfield  {journal} {\bibinfo
  {journal} {Phys. Rev. E}\ }\textbf {\bibinfo {volume} {89}},\ \bibinfo
  {pages} {062110} (\bibinfo {year} {2014}{\natexlab{b}})}\BibitemShut
  {NoStop}%
\bibitem [{\citenamefont {Torres-Herrera}\ and\ \citenamefont
  {Santos}(2014{\natexlab{c}})}]{Torres2014PRAb}%
  \BibitemOpen
  \bibfield  {author} {\bibinfo {author} {\bibfnamefont {E.~J.}\ \bibnamefont
  {Torres-Herrera}}\ and\ \bibinfo {author} {\bibfnamefont {Lea~F.}\
  \bibnamefont {Santos}},\ }\bibfield  {title} {\enquote {\bibinfo {title}
  {Nonexponential fidelity decay in isolated interacting quantum systems},}\
  }\href {\doibase 10.1103/PhysRevA.90.033623} {\bibfield  {journal} {\bibinfo
  {journal} {Phys. Rev. A}\ }\textbf {\bibinfo {volume} {90}},\ \bibinfo
  {pages} {033623} (\bibinfo {year} {2014}{\natexlab{c}})}\BibitemShut
  {NoStop}%
\bibitem [{\citenamefont {Santos}\ \emph
  {et~al.}(2012{\natexlab{a}})\citenamefont {Santos}, \citenamefont
  {Borgonovi},\ and\ \citenamefont {Izrailev}}]{Santos2012PRL}%
  \BibitemOpen
  \bibfield  {author} {\bibinfo {author} {\bibfnamefont {L.~F.}\ \bibnamefont
  {Santos}}, \bibinfo {author} {\bibfnamefont {F.}~\bibnamefont {Borgonovi}}, \
  and\ \bibinfo {author} {\bibfnamefont {F.~M.}\ \bibnamefont {Izrailev}},\
  }\bibfield  {title} {\enquote {\bibinfo {title} {Chaos and statistical
  relaxation in quantum systems of interacting particles},}\ }\href {\doibase
  10.1103/PhysRevLett.108.094102} {\bibfield  {journal} {\bibinfo  {journal}
  {Phys. Rev. Lett.}\ }\textbf {\bibinfo {volume} {108}},\ \bibinfo {pages}
  {094102} (\bibinfo {year} {2012}{\natexlab{a}})}\BibitemShut {NoStop}%
\bibitem [{\citenamefont {Santos}\ \emph
  {et~al.}(2012{\natexlab{b}})\citenamefont {Santos}, \citenamefont
  {Borgonovi},\ and\ \citenamefont {Izrailev}}]{Santos2012PRE}%
  \BibitemOpen
  \bibfield  {author} {\bibinfo {author} {\bibfnamefont {L.~F.}\ \bibnamefont
  {Santos}}, \bibinfo {author} {\bibfnamefont {F.}~\bibnamefont {Borgonovi}}, \
  and\ \bibinfo {author} {\bibfnamefont {F.~M.}\ \bibnamefont {Izrailev}},\
  }\bibfield  {title} {\enquote {\bibinfo {title} {Onset of chaos and
  relaxation in isolated systems of interacting spins: Energy shell
  approach},}\ }\href {\doibase 10.1103/PhysRevE.85.036209} {\bibfield
  {journal} {\bibinfo  {journal} {Phys. Rev. E}\ }\textbf {\bibinfo {volume}
  {85}},\ \bibinfo {pages} {036209} (\bibinfo {year}
  {2012}{\natexlab{b}})}\BibitemShut {NoStop}%
\bibitem [{\citenamefont {French}\ and\ \citenamefont
  {Wong}(1970)}]{French1970}%
  \BibitemOpen
  \bibfield  {author} {\bibinfo {author} {\bibfnamefont {J.B.}\ \bibnamefont
  {French}}\ and\ \bibinfo {author} {\bibfnamefont {S.S.M.}\ \bibnamefont
  {Wong}},\ }\bibfield  {title} {\enquote {\bibinfo {title} {Validity of random
  matrix theories for many-particle systems},}\ }\href {\doibase
  https://doi.org/10.1016/0370-2693(70)90213-3} {\bibfield  {journal} {\bibinfo
   {journal} {Physics Letters B}\ }\textbf {\bibinfo {volume} {33}},\ \bibinfo
  {pages} {449} (\bibinfo {year} {1970})}\BibitemShut {NoStop}%
\end{thebibliography}%

\end{document}